\documentclass[letterpaper, twocolumn]{article}

\usepackage[T1]{fontenc}

\usepackage{geometry}
\usepackage{setspace}

\usepackage[style=chem-acs, doi=true]{biblatex}
\usepackage{graphicx}
\usepackage{float}
\newfloat{scheme}{htbp}{los}
\floatname{scheme}{Scheme}
\floatname{chart}{Chart}
\newfloat{graph}{htbp}{loh}

\usepackage{hyperref} 
\usepackage{textgreek} 
\usepackage{caption}
\usepackage{crossrefs}
\SetupCrossRefs{main}
\input{si.labels}

\newcommand*\Vo{V\textsuperscript{4+}} \ignorespacesafterend
\newcommand*\Vp{V\textsuperscript{5+}}
\newcommand*\Vn{V\textsuperscript{3+}}
\newcommand*\Vnuc{\textsuperscript{51}V}
\newcommand*\Er{Er\textsuperscript{3+}}
\newcommand*\Otwo{O\textsubscript{2}}
\newcommand*\Ntwo{N\textsubscript{2}}

\newcommand*\TiOtwo{Ti\Otwo}

\newcommand*\thSiC{4\textit{H}-SiC}
\newcommand*\hhSiC{6\textit{H}-SiC}

\newcommand*\thk{4\textit{H}-\textit{k}}
\newcommand*\thh{4\textit{H}-\textit{h}}

\newcommand*\degC{\textdegree C}
\newcommand*\us{\textmu s}
\newcommand*\um{\textmu m}
\newcommand*\uW{\textmu W}

\newcommand*\cmnthree{cm\textsuperscript{-3}}

\usepackage{xcolor} 

\RenewDocumentEnvironment{abstract}{}{%
  \begin{center}\bfseries\Large\abstractname\end{center}%
  \begin{quote}
}{%
  \end{quote}%
}
\makeatletter
\if@twocolumn\else
  \renewcommand{\twocolumn}[1][]{#1}
\fi
\makeatother

\usepackage{authblk}

\author[1]{Timothy Draher\textsuperscript{\textdagger}}
\author[2,3]{Nolan Bitner\textsuperscript{\textdagger}}
\author[1]{Vasileios Niaouris\textsuperscript{\textdagger}}
\author[2,3]{Claire E. McDermott}
\author[4]{David Czaplewski}
\author[1,2,3]{Supratik Guha}
\author[1,2,3]{David D. Awschalom}
\author[1,2,3]{F. Joseph Heremans}
\author[1]{Alan M. Dibos*}

\affil[1]{Q-NEXT, Argonne National Laboratory, Lemont, IL 60439, USA}
\affil[2]{Pritzker School of Molecular Engineering, University of Chicago, Chicago Illinois 60637, USA}
\affil[3]{Material Science Division, Argonne National Laboratory, Lemont, Illinois 60439, USA}
\affil[4]{Center for Nanoscale Materials, Argonne National Laboratory, Lemont, Illinois 60439, USA}

\title{A Si-on-SiC Platform for Interfacing with Vanadium Dopants in the Telecom O-Band}

\date{\textsuperscript{\textdagger} Authors contributed equally \\ {*} Corresponding author: adibos@anl.gov}

\begin{document}

\twocolumn[
    \maketitle
    \begin{abstract}
        Vanadium color centers in silicon carbide (SiC) offer telecom O-band emission, sub-microsecond optical lifetimes, and second-scale spin relaxation times, rendering them promising candidates for quantum network nodes. 
        However, efficient photon extraction from the high-refractive-index SiC host remains challenging. 
        Here, we introduce a silicon-on-SiC heterogeneous photonic platform in which silicon nanocavities evanescently couple to shallow vanadium dopants in commercial \thSiC, enabling efficient zero-phonon line (ZPL) collection under resonant excitation. 
        In as-grown ensembles, we observe Purcell-enhanced photoluminescence with transient spectral hole burning linewidths of 50\,MHz on microsecond timescales. 
        In dilute implanted samples, we isolate individual vanadium centers with high-purity single photon emission and a Purcell-enhanced lifetime of 109\,ns. 
        The entire optical interface operates through a single lensed fiber, including a 905\,nm repump laser that recovers the vanadium charge state with 95\,\% efficiency.
        These results establish Si-on-SiC as a scalable, foundry-compatible platform for telecom-wavelength spin-photon interfaces.
    \end{abstract}
]


\vspace{5 pt}
\noindent \textbf{Abbreviations:} SiC, \thSiC, \Vo, Si, SOI, SiCOI, PLE, ZPL, PSB, TSHB

\vspace{10 pt}
Spin-photon interfaces with long-lived spin coherence and fast, spectrally stable optical transitions at wavelengths compatible with existing telecom fiber infrastructure offer a promising route toward building quantum networks at scale.
Several solid-state platforms have made significant progress toward this goal, including nitrogen-vacancy and group-IV vacancy centers in diamond~\cite{pompili2021rmq, knaut2024enq}, vacancy-based defects in silicon carbide~\cite{fang2024egs, babin2022fnw, zeledon2026mlq, lukin2023tem}, rare-earth ions in oxides~\cite{raha2020oqn, zhong2017nre, uysal2025spe, ruskuc2025mem}, and dopants/color centers in silicon~\cite{beregon2020sit,hollenbach2020ets,simmons2024sft}.
However, each of these platforms is limited by one or more challenges such as: non-telecom emission wavelengths, long radiative lifetimes, low emission efficiency in the zero phonon line (ZPL), spectral instability, or compatibility barriers with foundry-scale semiconductor fabrication and device integration.

\begin{figure*}[ht!]
    \centering
    \includegraphics[width=\linewidth]{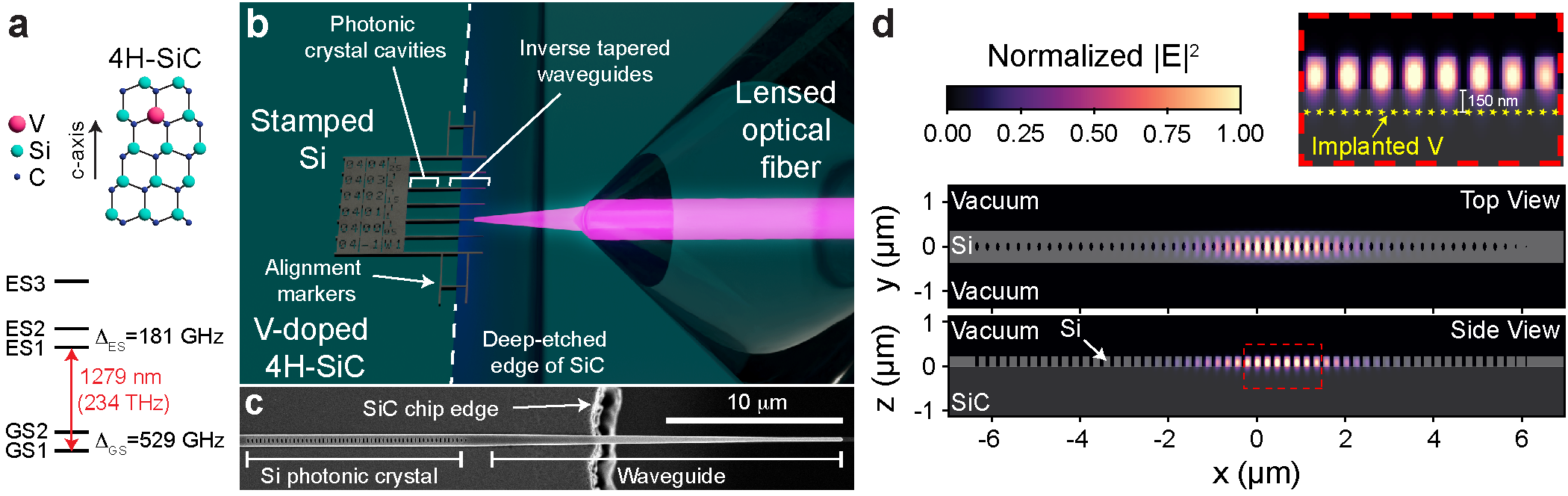}
    
    \caption{
    \label{fig1:V-doped_Platform}
    \textbf{(a)}\;(Top) Diagram of a vanadium dopant in the \thSiC\ lattice in a quasi-cubic \textit{k}-site. (Bottom) Simplified energy level diagram for \thSiC\ \textit{k}-site without an applied magnetic field. 
    \textbf{(b)}\;Schematic for lensed optical fiber coupling of Si photonic cavities stamped onto the edge of a pedestal etched into the SiC wafer. 
    \textbf{(c)}\;Scanning electron microscope image of a stamped nanocavity over the SiC pedestal edge. 
    \textbf{(d)}\;Horizontal (middle) and vertical (bottom) FDTD simulation cross-sections of the normalized electric field intensity $\|E\|^2$ of the TE-like mode along the center of the 220\,nm-thick and 700\,nm-wide Si waveguide. The dashed red box (top) highlights the evanescent field intensity overlap with implanted V at a simulated average depth of 150\,nm in Sample~B.
    }
\end{figure*}

Neutrally charged vanadium (\Vo) color centers in silicon carbide (SiC) have recently emerged as a competitive alternative~\cite{spindlberger2019opv,wolfowicz2020vsq,cilibrizzi2023uni, ahn2024esr, astner2024vsc}. 
SiC is an attractive, mature, foundry-scalable, and commercially available host material for optically active dopants.
Vanadium itself has been employed in SiC for decades as a deep-level compensation dopant in RF and high-power semiconductor devices~\cite{hobgood1995sis}, and as a result, V-doped semi-insulating SiC wafers are readily available from commercial vendors.
Vanadium dopants substitute for Si atoms, exhibiting native telecom O-band emission ($1278-1388$\,nm) due to transitions between the $3d$ orbitals of the dopant atom with lifetimes of $11-167$\,ns (with both wavelength and lifetime depending on the substitutional site in the common 4\textit{H}- and \hhSiC\ polytypes)~\cite{wolfowicz2020vsq, spindlberger2019opv}.
Operating in the O-band enables low-loss fiber links without the need for quantum frequency conversion and allows low-photon-number quantum signals to coexist with strong classical C-band traffic in the same fiber, mitigating noise from spontaneous Raman scattering~\cite{thomas2024qtc}.
At the same time, the \Vo\ features an optically addressable spin-1/2 electron ground state with spin relaxation times up to 28\,s at millikelvin temperatures, showing great potential for long electron-spin coherence, in addition to an intrinsic nuclear spin for \Vnuc\ ($I=7/2$) that could serve as a local quantum register~\cite{cilibrizzi2023uni, ahn2024esr, koller2025sec}.

Despite these favorable properties, efficient extraction of coherent photons from \Vo:SiC centers remains a central challenge. 
The relatively high refractive index of SiC ($n\approx2.6$) severely limits the fraction of emission collected by confocal microscopy, while the modest Debye-Waller factor of $0.25-0.5$ and theoretically estimated quantum efficiency of $\sim 2-20$\,\% further constrains usable photon rates \cite{wolfowicz2020vsq, spindlberger2019opv}. This has restricted all previous measurement of single \Vo:SiC defects to collection of photons from the incoherent phonon sideband (PSB) \cite{wolfowicz2020vsq, cilibrizzi2023uni}.
Nanophotonic integration is therefore essential for practical, coherent spin-photon interfaces. 
Monolithic approaches to SiC nanophotonics, including photoelectrochemical (PEC) etching of doped layers \cite{crook2020pes, xia2025ssc}, ion-beam angled etching \cite{Majety2025}, and SiC-on-insulator (SiCOI) substrates, are currently under development and promise to enable high quality-factor resonators and color-center integration \cite{yi2020wsc, Ou2024, bader2024arc}. 
Monolithic integration can maximize mode overlap between the defect and cavity, but simultaneously necessitates bringing the color centers into proximity with heavily etched surfaces which may degrade optical or spin coherence. 
Additionally, monolithic approaches add fabrication constraints, such as: PEC-based SiC nanophotonic devices require carefully structured layer doping profiles with etch stop layers to achieve the desired photonic layer thickness \cite{crook2020pes, xia2025ssc, Lipton2025}, and SiCOI devices must have transition metal dopants incorporated in them prior to the ion cut process because dopant activation in SiC requires temperatures sufficient \cite{wolfowicz2020vsq, cilibrizzi2023uni} to cause oxide layer reflow.

An alternative strategy is heterogeneous integration, in which emitters are evanescently coupled to photonic structures fabricated in a separate, mature, low-loss material. 
This approach has proven effective for rare-earth ions, where \Er\ emitters in YSO have been integrated with Si nanophotonics via stamping \cite{dibos2018ass} or \TiOtwo\ via thin-film deposition \cite{dibos2022pee}, as well as for vacancy-centers in bulk diamond coupled to stamp-transfered GaP films \cite{chakravarthi2023hig, yama2026sgp}. 
There has also been recent progress on the integration of thin-film Si and SiC with subsequent photonic device etching through the Si device layer \cite{devault2024ssc}.

In this Letter, we introduce a silicon-on-silicon-carbide (Si-on-SiC) heterogeneous photonic platform for telecom-wavelength \Vo\ color centers. 
By stamp-transferring Si nanobeam cavities onto SiC substrates, we evanescently couple near-surface vanadium dopants to Si  photonics. 
Initially, using as-grown \Vo\ ensembles, we observe Purcell-enhanced photoluminescence while retaining modest homogeneous linewidths compared to the lifetime-limit.
Then, by reducing the dopant concentration, we successfully isolate and resonantly address individual implanted \Vo\ centers, demonstrating enhanced efficiency of single-photon collection directly on the ZPL. 
We achieve these by implementing an in-fiber charge-stabilization protocol using a repump laser wavelength (905\,nm) slightly above the bandgap of Si. 
This repump restored the desired charge state with high efficacy while significantly reducing the parasitic absorption and heating associated with conventional higher energy visible or ultraviolet repump wavelengths \cite{wolfowicz2020vsq, ahn2024esr, cilibrizzi2023uni}.

Our efforts focus on the \thk\ site, which combines the longest reported spin relaxation lifetime among \Vo\ sites \cite{astner2024vsc, ahn2024esr} and the most commercially employed SiC polytype (Figure~\ref{fig1:V-doped_Platform}a; top).
For this site, the 3\textit{d}\textsuperscript{1} electron manifold splits in five distinct orbital states (two ground and three excited: GS and ES) due to the crystal-field and spin-orbit interactions, yielding a dominant GS1-ES1 ZPL at 1278.82\,nm with a lifetime of 167\,ns (Figure~\ref{fig1:V-doped_Platform}a; bottom).

To enhance light-matter interaction, we first define the Si photonic crystal nanobeam cavities via a lattice constant taper of elliptical holes in a standard 220\,nm-thick Si device layer silicon-on-insulator (SOI) substrate and stamp-transfer them onto the edge of deep-etched (100\,\um) SiC pedestals (Figure~\ref{fig1:V-doped_Platform}b-c, and Supporting Information (SI), Section~\ref{SIsec:fab}).
The nanocavity is designed for a fundamental transverse-electric (TE)-like mode with an evanescent field intensity ($\left|E\right|^2$) with 1/$e$ decay length of 75\,nm into the SiC (Figure~\ref{fig1:V-doped_Platform}d). 
Our initial design yields maximum FDTD-simulated cavity quality factor of $Q^\mathrm{max}_\mathrm{ideal} = 1.4 \times 10^{5}$ (see SI, Figure~\ref{figS:sim_scale_sweep}). 
However, as-fabricated devices show a consistent geometric distortion of the small fill fraction cavity holes; simulations that use these extracted geometric parameters from SEM images yield a lower expected maximum cavity Q factor of $Q^\mathrm{max}_\mathrm{distorted} \approx 7 \times 10^{4}$ (see SI, Figure~\ref{figS:tuning_taperCoupling}a). 
For the type of cavities used in this work (4 mirror holes pairs, $Q_\mathrm{distorted} \approx 3.5 \times 10^{4}$) the maximum simulated Purcell factor for an emitter in SiC will occur at the Si/SiC interface to yield $F^\mathrm{max}_{P}\approx 980$. 
This drops to $F_{P}\approx 135$ at a depth of 150\,nm into the SiC, at the center of the cavity.

\begin{figure*}[!ht]
    \centering
    \includegraphics[width=\linewidth]{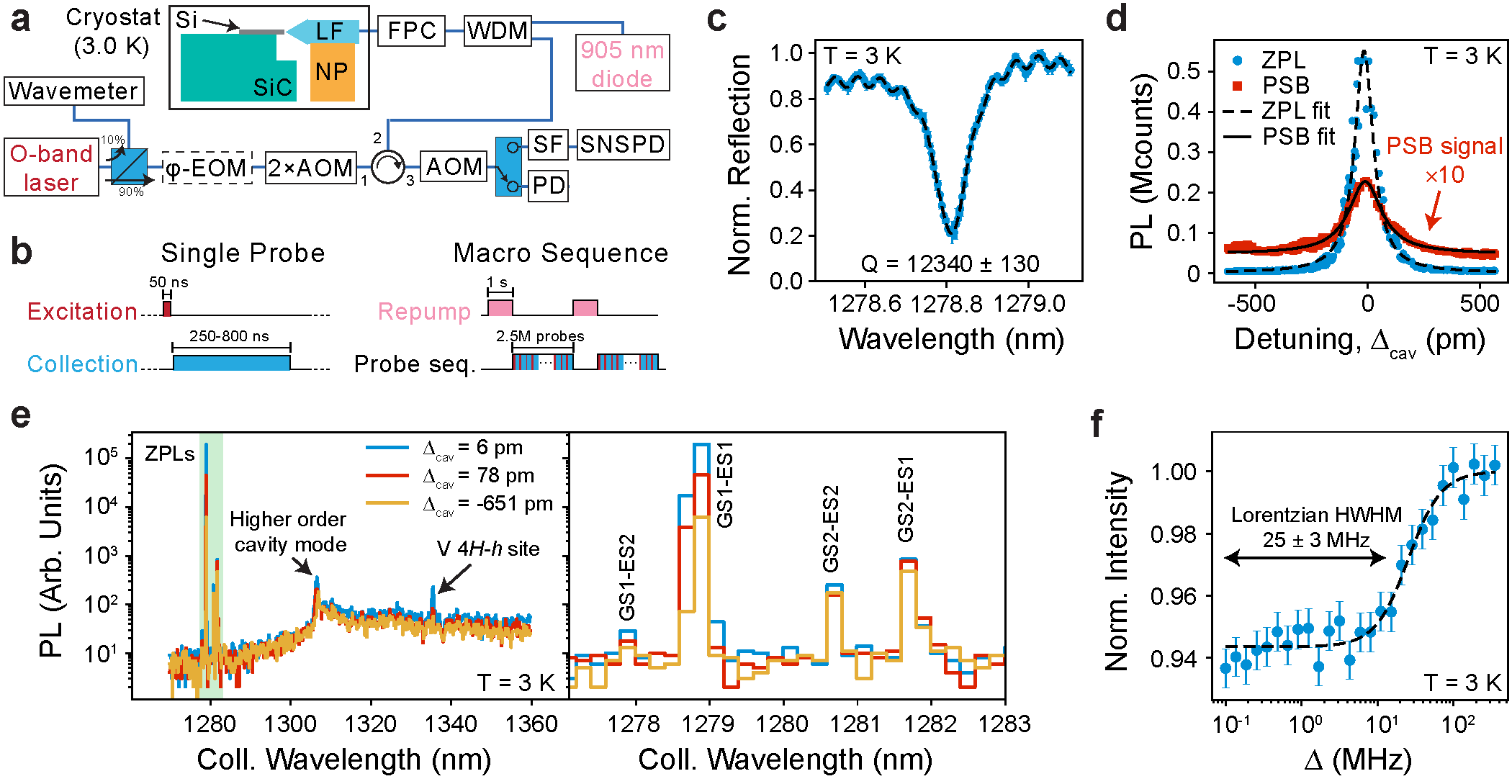}
    
    \caption{
    \label{fig2:bulk}
    \textbf{(a)}\;Simplified experimental configuration utilizing a fiber circulator. A tunable O-band laser in combination with two acousto-optic modulators (AOM) produces short pulses of light. A lensed fiber (LF) on a nanopositioner (NP) couples light to and from the nanocavity inside a cryostat. A wavelength division multiplexer (WDM) is used to inject 905 nm light for V dopant charge state repumping, and a fiber polarization controller (FPC) is used to align the polarization to the nanocavity. 
    A switch directs photons to an InGaAs photodiode (PD) for reflection monitoring or a superconducting nanowire single-photon detector (SNSPD) after a variable spectral filter (SF) chosen for the experiment. An AOM in the collection path prevents SNSPD saturation via resonant laser pulses. 
    \textbf{(b)}\; Individual laser pulse sequence used for resonant excitation and photoluminescence (PL) collection, and the associated macro experimental sequence whereby 905\,nm light charge repump pulses are used before millions of successive single probe cycles.
    \textbf{(c)}\; Normalized Si-on-SiC nanocavity laser reflection spectrum on 4\,ppm bulk V-doped (Sample~A) tuned near the GS1-ES1 optical transition ($\lambda=1278.82$\,nm). The spectrum is fit to extract the cavity quality factor (Q).
    \textbf{(d)}\; The integrated PL signal of V ensemble in cavity from (c) as a function of laser-cavity detuning ($\Delta_{cav}$) when pumping at GS1-ES1 and collecting fluorescence in two different spectral ranges: zero phonon line (ZPL) bandpass ($1250-1300$\,nm) or phonon sideband (PSB) longpass ($>1300$\,nm).
    \textbf{(e)}\;(Left) PL spectrum of the same cavity-coupled V ensemble in (d) that has been excited at the GS1-ES1 ZPL for three different values of $\Delta_{cav}$. A tunable filter (FWHM = 100\,pm) in the collection path spectrally resolves the four ZPLs (green box) and the broad PSB. (Right) A zoom-in of the green ZPL region. 
    \textbf{(f)}\; Transient spectral hole burning spectrum of V ensemble in (d)-(e) resonant with the nanocavity, showing the normalized PL intensity as a function the carrier-sideband detuning ($\Delta$) set by the frequency on the phase-EOM ($\phi$-EOM, in (a)) with a Lorentzian fit of the half-width half-maximum (HWHM).
    }
\end{figure*}

To characterize our devices and the coupled \Vo\ properties, we employ a fully fiber-coupled optical setup that exploits the platform’s native compatibility with single-mode fiber networks, bypassing the need for free-space optics or microscope objectives (Figure~\ref{fig2:bulk}a). 
Our deliberate choice to employ a lensed optical fiber and the inverse tapered Si waveguide is to enable high efficiency coupling with broadband optical excitation: 
(i) O-band laser excitation resonant with the GS1-ES1 ZPL transition, and 
(ii) a second charge repump wavelength far detuned from the ZPL, which would be challenging to achieve with narrowband, high-efficiency grating couplers \cite{Tao2025}. 
Our hybrid device design enables relatively high efficiency extraction of cavity-collected photons directly to optical fiber, with simulated fiber-to-waveguide coupling efficiency of 63\,\% (see SI, Figure~\ref{figS:tuning_taperCoupling}d), which can lead to substantial improvements in the overall photon collection rates compared to bulk. 
After pulsed resonant excitation (with AOM rise/fall times, $\tau_{R} = 15-20$\,ns), we isolate specific photoluminescence (PL) emission channels and suppress background counts through the use of interchangeable spectral filters connected to the optical switch, including a ZPL bandpass filter ($1250-1300$\,nm), a PSB longpass filter ($>1300$\,nm), or a tunable narrowband electronic filter (see SI Section~\ref{SIsubsec:config}).

The temporal sequences employed for time-resolved PL collection (Figure~\ref{fig2:bulk}b) are designed to maximize the charge stability of the \Vo\ centers and the PL signal-to-noise ratio, as optimized on a single emitter (discussed below). 
Each measurement run begins with a 905\,nm charge-repump pulse to prepare the dopants in the desired \Vo\ charge state, followed by millions of successive resonant O-band excitation probe pulses. 
During the excitation probe pulse, the collection AOM is closed to shield the SNSPD from the strong reflected laser pulse. 
When the probe pulse turns off, we wait for $2 \tau_{R}$ before we open the collection path AOM, which eliminates laser bleed-through.

In this work we explore coupled \Vo\ dopants via two \thSiC\ samples with different doping profiles (see SI, Table~\ref{tab:samples} for summary). 
Sample~A is an as-grown, V-doped, semi-insulating, \thSiC\ substrate with a uniform doping density of approximately $1.9\times10^{17}$\,\cmnthree.
Prior to cavity stamping, the only process Sample~A's SiC substrate undergoes is the deep-etched pedestal fabrication (see SI, Section~\ref{SIsec:fab}). 
Sample~B, in contrast, is a high-purity semi-insulating (HPSI) \thSiC\ substrate that has undergone vanadium ion implantation at a fluence of $1\times10^{11}$ cm$^{-2}$, energy of 275\,keV (mean depth of 150\,nm), and temperature of 500\,\degC. 
We annealed Sample~B at 1400\,\degC\ in an argon atmosphere at ambient pressure to incorporate the dopants into Si-substitutional sites (see SI, Section~\ref{SIsec:fab}). 
We use carbon capping to reduce roughness induced by high-temperature annealing~\cite{capano1999sri, keubler2024rcc}, and measure surface roughness values in the range of $0.7-2$\,nm RMS. 
A key technical challenge in our heterogeneous approach is achieving smooth \thSiC\ and gap-free contact between the Si nanocavities and the top surface of the SiC, which otherwise leads to resonance shifting, excess scattering losses, and poorer cavity-to-waveguide coupling efficiencies. 
It is important to note that Sample~A consists of devices with higher Q factors, but lower waveguide-to-fiber coupling efficiencies as a result of a Si inverse taper design error. 
In contrast, Sample~B consists of devices with slightly lower Q factors due to increased mean SiC surface roughness after annealing, but greatly improved waveguide-to-fiber coupling of 67\,\%, slightly exceeding the 63\,\% simulated value.

For both samples, we target nanocavities that are near 1279\,nm at 3\,K ($\sim1289$\,nm at room temperature; see SI, Figure~\ref{figS:tuning_taperCoupling}c). 
A typical normalized reflection spectrum at 3\,K is shown in Figure~\ref{fig2:bulk}c, demonstrating a well-defined fundamental cavity mode at $\lambda_{cav}=1278.82$\,nm. 
We perform a Lorentzian-based fit (see SI, Section~\ref{SIsec:Design}) and measure a cavity full-width half-maximum (FWHM) of $103.6\pm 1.1$\,pm ($19.0 \pm 0.2$\,GHz) for a quality factor of $12\times 10^3$ and 
cavity-to-waveguide coupling efficiency ($\eta_\mathrm{CW}$) of 27\,\% (undercoupled regime). 
Precise spectral alignment with the \Vo\ ZPL at cryogenic temperatures within a few nm (see SI, Figure~\ref{figS:tuning_taperCoupling}) requires deterministic tuning via \Ntwo\ condensation \cite{mosor2005spc, dibos2018ass}.

Using Sample~A, we perform three complimentary sets of PL experiments to look at optical performance: 
(i) cavity-mediated emission enhancement while pumping on the GS1-ES1 ZPL as a function of cavity-laser detuning, 
(ii) spectrally-selective spectroscopy of the collected emission while pumping on the GS1-ES1 ZPL as a function of cavity-laser detuning, and 
(iii) transient spectral hole burning experiments with the cavity tuned to the GS1-ES1 ZPL as a proxy for homogeneous optical linewidth. 
The relatively large number of \Vo\ in this sample enables sufficient signal-to-noise for evaluation of general emission behavior in the presence of the cavity.

We first quantify the cavity-mediated emission enhancement by slowly sweeping the cavity resonance across the GS1-ES1 transition while monitoring the collected photon counts under resonant excitation (Figure~\ref{fig2:bulk}d). 
By alternating between the ZPL and PSB filters, we independently track the cavity’s effect on each emission channel.
When the cavity is far-detuned from the GS1-ES1 transition, the excitation pulse only addresses bulk emitters coupled directly underneath the waveguide, as the emitters underneath the cavity region are not efficiently excited as the laser wavelength is in the photonic stopband.
When the cavity is tuned onto resonance, the ZPL emission increases by up to $121\times$ relative to the far-detuned background, with an enhancement linewidth matching the cavity (100\,pm). 
In contrast, the PSB emission shows only a $4.1\times$ enhancement with a broader linewidth of 178\,pm. 
Concurrently, the radiative lifetime measured on resonance decreases from the bulk value of 167\,ns \cite{wolfowicz2020vsq} for ZPL-filtered emission (stretched-exponential time constant, $\tau$) to 121\,ns, and to 140\,ns for PSB-filtered emission (see SI, Section~\ref{SIsec:ensemble-lifetimes}). 
Because these population increase and rate enhancement contributions cannot be fully disentangled in an ensemble measurement, we estimate an upper bound on the Purcell factor by taking the ratio of the ZPL to PSB count-rate increases (and correcting for background counts), $F_P \leq 120 / 3.1 \simeq 38$, under the simplistic assumption that both channels experience the same increase in excited population.

To spectrally resolve the cavity-enhanced emission, we scan a tunable narrowband filter (100\,pm width and step size) across the collection path from 1270\,nm to 1360\,nm while resonantly exciting the GS1-ES1 transition (Figure~\ref{fig2:bulk}e; see SI Section~\ref{SIsubsec:ensemble-coll-dep}). 
This range encompasses four ZPL transitions as well as part of the PSB \cite{spindlberger2019opv, wolfowicz2020vsq}. 
At three representative cavity detunings, we observe the expected four ZPL transitions. 
With the cavity far-detuned, the GS1-ES1 line accounts for 90\,\% of the total ZPL emission; on resonance, the cavity funnels 99.4\,\% of ZPL emission into this single line (see SI, Section~\ref{SIsubsec:ensemble-coll-dep}).
A higher-order cavity mode and the \thh\ site are also visible in this spectrum (see SI, Section~\ref{SIsubsec:ensemble-coll-dep}). 

Having characterized the cavity-enhanced emission spectrum, we probe the optical coherence of the coupled ensemble via transient spectral hole burning (TSHB) (Figure~\ref{fig2:bulk}f) \cite{Weiss21}.
Under low-power conditions, we resolve a minimum a half-width at half-maximum (HWHM) of $25\pm3$\,MHz, corresponding to a homogeneous linewidth of 50\,MHz (Figure~\ref{fig2:bulk}d), with no indication that this is the lowest achievable linewidth (see SI Section~\ref{SIsec:TSHB}, Figure~\ref{figS:tshb-vs-power}).
This homogeneous linewidth is much narrower than the cavity linewidth, placing this system in the bad cavity limit. 
However, it remains much broader than the 1.3\,MHz transform limit ($1/(2\pi\tau)$ for $\tau=121$\,ns; see SI, Figure~\ref{figS:lifetime-summary}) of the cavity-enhanced emission on resonance. 
This indicates that even on the 60\,\us\ timescale of our TSHB sweep time window (see SI Section~\ref{SIsec:TSHB}), the emitters may be subject to spectral diffusion driven by slow fluctuations in the local charge environment.

\begin{figure}[!ht]
    \centering
    \includegraphics[width=0.9\linewidth]{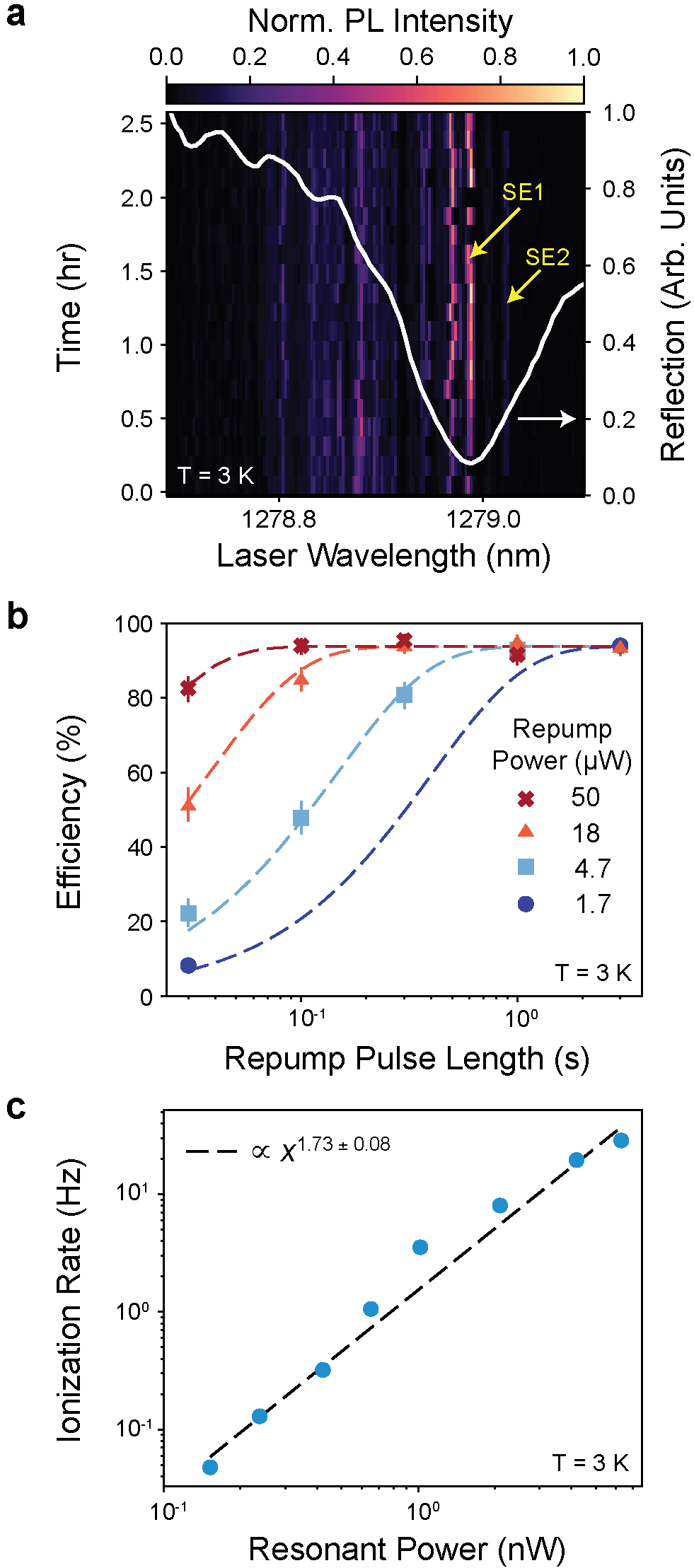}
    \caption{
    \textbf{(a)} Time-dependent photoluminescence excitation (PLE) spectroscopy of implanted V dopants coupled to a cavity. The normalized reflection spectrum of the cavity is overlaid in white (right axis), and has been tuned to overlap with an isolated spectral peak (SE1) at 1278.99\,nm.
    \textbf{(b)} 905\,nm repump efficiency as a function of pulse duration and power. The fit lines are a single exponential to the whole dataset using total pulse energy as the only independent variable. The repump efficiency is observed to saturate at 95\,\%.
    \textbf{(c)} Measured ionization rate of SE1 as a function of the resonant laser power. The dashed line is a power law fit to the ionization rate showing superlinear behavior.
    }
    \label{fig3:charge}
\end{figure}

To search for single-emitter candidates, we perform time-resolved photoluminescence excitation (PLE) scans across the inhomogeneous distribution of V-implanted Sample~B. 
We perform successive PLE scans where we use a charge-repump sequence on every pixel (Figure~\ref{fig3:charge}a). 
Near the GS1-ES1 line center, the spectral density of emitters remains too high to isolate individual emitters. 
However, on the red edge of the distribution, we identify two spectrally isolated and stable peaks (SE1 and SE2) that persist across successive scans, while the cavity is resonant with SE1 ($Q\approx7.9 \times 10^3$ and $\eta_\mathrm{CW}\approx35$\,\%). 

Without periodic charge repumping, SE1 turns off entirely (see SI Section~\ref{SIsubsec:no_repump}, Figure~\ref{figS:no_repump}). 
This is a result of photoinization to either the dopant's donor (\Vp) or acceptor (\Vn) charge state under resonant excitation~\cite{ahn2024esr, cilibrizzi2023uni}. 
Both the \Vp\ and \Vn\ charge states are optically dark under O-band illumination and previous work has not been able to conclusively identify which charge state is the primary dark state after photoionization, so short wavelength lasers that can re-ionize both dark charge states are typically used for repumping.

We use a 905\,nm diode laser delivered through the same fiber as the resonant light via the WDM to recover the \Vo\ charge state. 
This wavelength is chosen to lie closer to the Si bandgap at 3\,K (1.17\,eV, 1060\,nm) compared to the conventional visible or UV repump sources employed in prior work \cite{wolfowicz2020vsq, ahn2024esr, cilibrizzi2023uni}, which substantially reduces parasitic absorption by the cavity and resultant cavity resonance drift.

To quantify the repump efficiency, we employ a deterministic ionization-recovery protocol which entails strong resonant excitation pulses to drive the defect into the dark state, followed by a 905\,nm repump pulse of variable length and power, and finally a charge-state check to determine the final charge state after the pulse (see SI, Section~\ref{SIsubsec:repump}). 
We repeat this many times to extract the recovery probability as the fraction of successful events (see SI, Figure~\ref{figS:repump-eff-table}b). 
The repump efficiency is well described by a single-exponential fit dependent on only the total pulse energy, and the maximum observed repump efficiency saturates at 95\,\%  (Figure ~\ref{fig3:charge}b). 
For measurements throughout this Letter, we use 1\,s repump pulses at modest power (4.7\,\uW), each pulse provides sufficient energy for near-complete recovery while allowing thermal dissipation over longer timescales that is compatible with cooling power constraints of typical dilution refrigerators at base temperature. 

We then characterize SE1's ionization dynamics under resonant excitation to identify the optimal probe conditions (see SI, Section~\ref{SIsubsec:resonant_pumping}). 
At a range of resonant excitation powers, we perform a sequence of repump-probe measurements and observe 
an exponential decay corresponding to photoionization of the \Vo\ center to the dark state, with a rate that increases with excitation power following a power-law dependence with exponent $1.73 \pm 0.08$ (Figure~\ref{fig3:charge}c; SI Figure~\ref{figS:ion-power-dep-raw}).
At a lower excitation power (240\,pW), the ionization rate drops to 0.13\,Hz maintaining an excellent signal-to-noise ratio, defining the optimal operating point for probing single emitters.

\begin{figure*}[!ht]
    \centering
    \includegraphics[width=\linewidth]{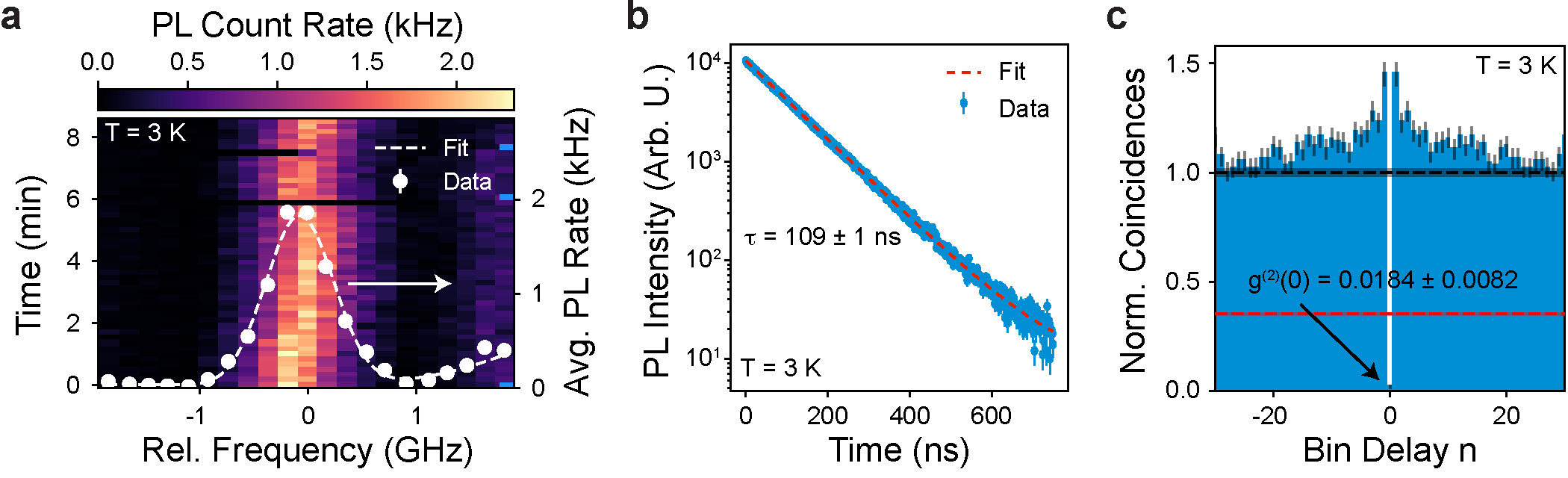}
    \caption{
    \textbf{(a)} (Left axis) PLE spectroscopy over time of a single emitter (SE1) resonant with the cavity with intensity displayed in units of photons detected per second of collection time. Blue marks near the right axis indicate instances when a 905\,nm repump pulse is applied between scans. (Right axis) An average of the intensity for the entire measurement duration and a Lorentzian fit are overlaid in white. 
    \textbf{(b)} Optical lifetime measurement of SE1 when on resonance with the cavity showing a reduced lifetime relative to bulk emitters due to Purcell enhancement.
    \textbf{(c)} Pulsed autocorrelation measurement of SE1 confirming that it is a single emitter (reflected about n=0 for clarity). Each bin corresponds to a pulse cycle of 992\,ns.
    }
    \label{fig4:single}
\end{figure*}

We use the optimized charge stabilization protocol to measure a time series of narrow-range scans near SE1 (Figure ~\ref{fig4:single}a), where the 905\,nm repump pulse is triggered only when the collected signal at the known emitter resonance drops below a fixed threshold at the beginning of each scan. 
While this requires an a priori estimate of the emission wavelength and expected count rate, it minimizes perturbation to the emitter and its local charge environment between successive wavelength steps and scans, reducing repump-induced spectral diffusion. 
Over the course of several minutes, we perform 12\,s scans requiring only two repump events, while the emission wavelength remains relatively stable.

The cumulative PLE data yield an emitter linewidth of 750\,MHz, which reduces to 675\,MHz when drift-free scans are selected (see SI, Figure~\ref{figS:no_repump}). 
Under resonant excitation, the lifetime of SE1 is 109\,ns (Figure~\ref{fig4:single}b), corresponding to a transform-limited linewidth of 1.45\,MHz. 
The measured linewidth exceeds this limit by more than two orders of magnitude, indicating that spectral diffusion on timescales up to seconds dominates the observed broadening. 
This is consistent with the ensemble TSHB measurement (Figure~\ref{fig2:bulk}f), which revealed a 50\,MHz homogeneous linewidth on the 60\,\us\ timescale of the frequency sweep -- substantially narrower than the single-emitter linewidth integrated over seconds, but still well above the transform limit, pointing to spectral diffusion across multiple timescales. 
Compared to the bulk lifetime of 167\,ns, the reduced lifetime of SE1 is consistent with Purcell enhancement of the ZPL emission rate by the cavity mode. 
The single-emitter lifetime avoids the averaging effects that broaden the decay in the ensemble measurement (121\,ns) despite a slightly higher cavity Q for Sample~A. 

Finally, we verify the single-photon nature of SE1 via a pulsed autocorrelation measurement (Figure~\ref{fig4:single}c) under resonant excitation and ZPL collection. We use a single SNSPD with a dead time of $\tau_d=40$\,ns, comparable to the emitter lifetime $\tau=109$\,ns, in this case the single emitter discrimination limit must be corrected to a stricter threshold of 0.346 as opposed to the conventional value of 0.5 (see SI, Section~\ref{SIsubsec:g2-theory}). 
Even against this reduced limit, we measure a non-background-corrected $g^{(2)}(0) = 0.0184 \pm 0.0082$ (Figure~\ref{fig4:single}c), consistent with the detector dark-count floor and unambiguously confirming SE1 as a single emitter. 
We attribute the bunching at small bin delays ($20>|n|\geq1$) to spectral diffusion based upon the dramatic reduction in bunching when SE1 is measured at increased temperature (6\,K; see SI, Figure~\ref{figS:g2-lifetime-2}a). 
We also obtain similar lifetime and antibunching results for isolated peak SE2 (see SI, Section~\ref{SIsubsec:add_g2}).

In summary, we introduce a Si-on-SiC heterogeneous photonic platform that addresses one of the primary challenges of realizing a telecom-wavelength quantum node with \Vo\ color centers - efficient extraction of ZPL photons into optical fiber.
By stamp-transferring Si nanobeam cavities onto commercial SiC substrates, we demonstrate Purcell-enhanced emission, resonant single-emitter addressing with all photons collected on the ZPL, and---based on the observed reduction of lifetime (from 167\,ns to 109\,ns)---dopant to fiber collection efficiencies of at least 21\,\% (see SI, Section \ref{SIsec:QE}). 
Future implementations of this platform can achieve even more impressive photonic enhancement and photon collection efficiency through relatively straightforward fabrication improvements.
The introduction of atomic layer etching (ALE) of the SiC surface before cavity stamping \cite{michaels2023bpa} and optimized annealing protocols, could reduce the surface roughness induced scattering, that currently limits the cavity quality factor here to $\sim 10^4$. Refinement of the cavity design can further increase $Q$ toward the peak simulated values near $10^5$. 
Furthermore, a reasonable reduction in the depth of the V dopants (or ALE to bring deeply embedded V-dopants closer to the surface) from 150\,nm to 15\,nm should achieve a roughly 20-fold increase the maximum Purcell factor.

However, our measurements clearly indicate narrowing the single-emitter optical linewidth toward the transform limit to enable indistinguishable photon generation for remote entanglement is a necessary challenge to address for this platform. 
Both the 675\,MHz single emitter linewidth measured over seconds, and the ensemble TSHB linewidth of 50\,MHz measured on the 60\,\us\ timescale---both well above the 1.45\,MHz transform limit---are indicative of significant spectral diffusion across multiple timescales.
We attribute much of this spectral diffusion to charge fluctuations of background impurities in the commercial substrates used in this work which can act as charge donors or acceptors, and have been seen to contribute to spectral diffusion for V2 color centers in similar grade HPSI \thSiC~\cite{vandeStolpe2025}.
Furthermore, the implantation process used to introduce dopants in Sample~B inevitably creates lattice damage and residual charge traps, which likely exacerbate the issue.
These factors indicate a set of improvements that could be made to address the excess spectral diffusion.
Replacing ion implantation with delta-doping during epitaxial SiC growth would simultaneously eliminate implantation and annealing damage, place emitters at a controlled shallow depth for stronger evanescent coupling to the cavity, and embed them in higher-purity material with fewer background charge traps. 
Furthermore, as demonstrated for other color centers in SiC~\cite{anderson2019eoc}, complementing this with on-chip electrical gates adjacent to the cavity would actively stabilize the residual charge environment. 
Importantly, the non-destructive nature of the heterogeneous integration preserves the crystal quality: the measured single-emitter linewidth of 675\,MHz agrees closely with values reported for implanted \Vo\ centers in bulk SiC~\cite{wolfowicz2020vsq, cilibrizzi2023uni}, confirming that the heterogeneous integration does not measurably degrade the optical coherence of the implanted emitters, and therefore improvements to the underlying SiC host should also be reflected in integrated devices.

The results presented here already satisfy most optical front-end requirements for a telecom O-band quantum network node that can co-exist with classical C-band traffic: resonant single-emitter addressing, ZPL single-photon collection, and fiber-coupled charge management. 
The platform also provides a strong foundation for pursuing the remaining milestones required for a high quality quantum node: coherent spin control (including of the intrinsic \Vnuc\ nuclear spin which offers a built-in quantum register for quantum error correction and long-term state storage~\cite{tissot2022nsq, koller2025sec}) via on-chip microwave delivery, sub-100\,mK operation with an applied magnetic field to resolve and address the spin-1/2 ground-state transitions, and suppression of spectral diffusion to enable two-photon interference between remote emitters. 
Finally, this platform provides clear advantages to future scaling with fiber-native operation enabled by charge management with near-Si-bandgap light and compatibility with commercially available material and foundry-compatible processes.

\section*{Acknowledgments}

The authors thank Suzanne Miller and Ralu Divan for nanofabrication tool assistance. The authors also thank Connor Horn for help with nickel electroplating. This work is primarily funded by the U.S. Department of Energy Office of Science National Quantum Information Science Research Centers as part of the Q-NEXT center, providing support for overall platform development and experiments, including simulations, device design, and fabrication capabilities including flip-chip bonding and high-temperature annealing (T.D., C.E.M., S.G., and A.M.D). The cryogenic optical measurements have additional support from the U.S. Department of Energy, Office of Science, Basic Energy Sciences, Materials Sciences and Engineering Division (N.B., D.D.A, and F.J.H.). Additional measurement infrastructure, experimental design, and data analysis were supported by the U.S. Department of Energy, Office of Science, Advanced Scientific Computing Research (ASCR) under Contract No. DE-AC02-06CH11357 as part of the InterQnet quantum networking project (V.N.). Device fabrication was performed at the Center for Nanoscale Materials, a U.S. Department of Energy Office of Science User Facility, was supported by the US DOE, Office of Basic Energy Sciences, under Contract No. DE-AC02-06CH11357.

\section*{Supporting information}

Additional fabrication, simulation, ensemble and single emitter characterization details are available in the Supporting Information.

\printbibliography

\end{document}


\twocolumn[\maketitle]
\clearpage
\tableofcontents
\clearpage

\renewcommand{\thefigure}{S\arabic{figure}}
\setcounter{figure}{0}

\section{Si-on-SiC Device Fabrication}
\label{SIsec:fab}

\begin{figure}[H]
    \centering
    \includegraphics[width=1\linewidth]{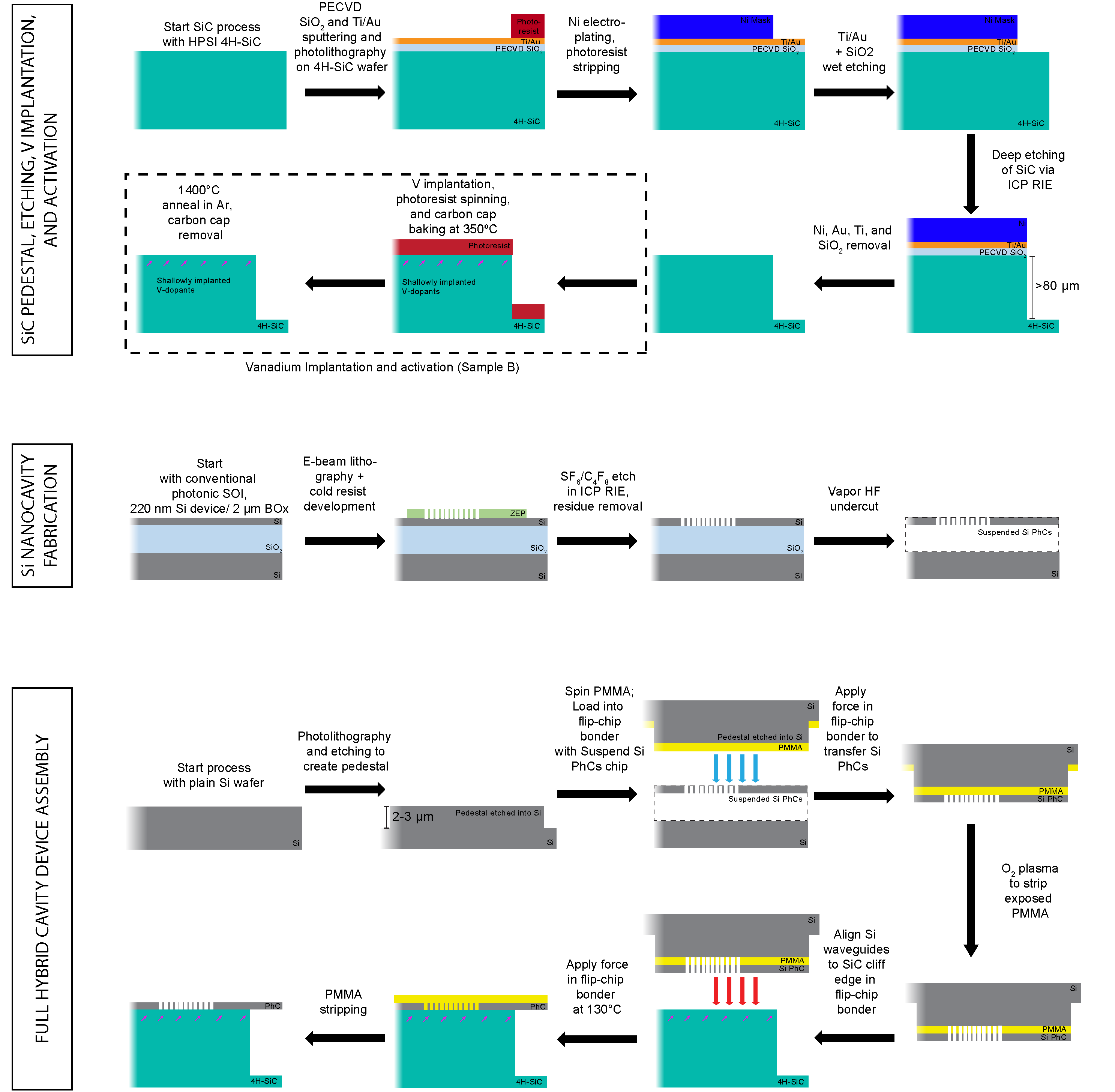}
    \caption{Full Si-on-SiC process flow including (1) tall SiC pedestal etching, (2) optional V implantation, and annealing, (3) suspended Si nanocavity fabrication, and (4) final hybrid device assembly.}
    \label{fig:dev_assembly}
\end{figure}

Figure~\ref{fig:dev_assembly} summarizes the complete process flow including the Si nanocavity fabrication, processing to create tall SiC pedestals, and flip-chip bonding to create finished Si-on-SiC devices. 

\subsection{SiC pedestal fabrication and V dopant implantation and activation}

\subsubsection{Fabricating Tall SiC Pedestals}

We fabricate tall pedestals on 4-inch \thSiC\ wafers to provide a flat, preserved stamping surface next to a deep, straight edge for lensed fiber access to our edge-coupled photonic crystal cavity devices. All processing is performed on the Si face of the wafer. 

First, we create shallow alignment markers via photolithography (Heidelberg MLA-150, Shipley S1805 resist) and transfer these $\sim100$\,nm into the SiC via \CHFthree/\SFsix\ ICP reactive ion etching. 
We then deposit a sacrificial protective silicon oxide layer ($\sim 40$\,nm \SiOtwo\ by Oxford PECVD at 300\,\degC) and an electroplating metal seed layer ($\sim5$\,nm Ti and 50\,nm Au via AJA sputtering). 
A thick negative photoresist layer (AZ nLOF 2070, 375\,nm exposure) is patterned such that the \emph{uncoated} Au areas enable Ni electroplating for protection of the SiC surface for subsequent deep etching.
We then electroplate a thick ($\sim 5$\,\um) Ni film into the bare Au to form an etching mask. 

To mitigate ragged edges after protection layer removal, the plated Ni thickness is kept below the resist height. 
After electroplating, we strip the AZ nLOF 2070 resist in solvent, and we subsequently remove the exposed Au, Ti, and \SiOtwo\ layers via wet etching while the Ni-covered pedestal tops remain intact. 
Deep etching of the SiC uses reactive ion etching (Oxford Cobra ICP) with \SFsix\ chemistry at high power recipe (2500\,W ICP, 300\,W RF). 
We etch in 15\,min segments with 3--5\,min cooling times to limit substrate heating and Ni erosion. A total etch time of $\sim2.5$\,hr yields $\sim 75-100$\,\um\ tall pedestals. 
Following the deep SiC etch, residual Ni and Au/Ti/\SiOtwo\ layers are stripped in respective wet etch steps. 
Critically, the \SiOtwo\ sacrificial layer remains protected under the various metal layers during the deep etch, which helps preserve a smooth pedestal edge for device stamping.

\subsubsection{Vanadium Implantation and Activation (Sample~B)}

This subsection describes the post-pedestal fabrication implantation and anneal process used for Sample~B (shown in in Figures 3-4 of the main text). 
In contrast, Sample~A (Figure~\ref{fig2:bulk}) uses as-grown V-doped ($\sim$ 4\,ppm) HPSI \thSiC\ and does not undergo these steps.

High-purity semi-insulating (HPSI) \thSiC\ substrates (Wolfspeed) are implanted with vanadium at a fluence of $1\times10^{11}$\,\cmntwo, an energy of 275\,keV, and a substrate temperature of 500\,\degC\ (CuttingEdge Ions), yielding a projected implant depth of $\sim 150$\,nm. 
Prior to the activation anneal, a carbon cap is formed by spinning a few microns of photoresist and baking at 350\,\degC\ in air to carbonize the film. 
This cap protects the SiC surface from carbon diffusion and graphitization~\cite{keubler2024rcc} during the subsequent high-temperature anneal. 
The sample is then annealed at 1400\,\degC\ in an Ar atmosphere (AnnealSys Zenith) at ambient pressure for 30 minutes to incorporate the vanadium into Si-substitutional lattice sites. 
After the high-temperature anneal, the carbon cap is removed by annealing at 750\,\degC\ in air, followed by a submersion in buffered HF.
The carbon capping layer reduces the degree of annealing induced roughness. However, post-anneal AFM measurements (Bruker FastScan) still show values of 1--2 \,nm RMS roughness (as compared to the as-received roughness values in the range of 0.2--0.3\,nm RMS). 
We believe this roughness limits the cavity quality factors on Sample~B to $\sim 8\times10^3$ due to scattering losses at the Si--SiC interface.

\subsection{Si nanocavity fabrication}

Silicon photonic crystal nanobeam cavities are patterned into 220\,nm thick Si device layer silicon-on-insulator (SOI) pieces with 2 microns of buried oxide (Soitec). 
We use e-beam lithography system (JEOL 8100FS) at a beam current of 2\,nA to pattern resist (undiluted ZEP 520A). 
After exposure, the resist is cold-developed in n-amyl acetate at 2\,\degC\ for 90\,s. 
The pattern is transferred in the Si device layer via plasma etching (\SFsix/\CfourFeight\ in PlasmaTherm ICP-RIE). 
We then strip the ZEP mask and fluorinated residue via heated solvent. 
We suspend the Si waveguide using a vapor hydrofluoric acid (HF) system (SPTS uEtch) to etch the buried oxide. 
The array of nanocavities are arranged into a "block" of typically 70-80 separate devices to release and transfer together. 
This is done to increase structural integrity during stamping and to increase the surface area for higher bonding yield. 
Each block contains various carefully engineered break points to remain suspended after the vapor HF undercut, but also enable facile transfer during stamping via subsequent flip-chip bonding steps.

\subsection{Full Device Assembly}

We use two-step flip-chip processes for final Si-on-SiC device assembly: (1) Si nanocavity pickup from the suspended Si device chip and (2) Si-on-SiC bonding via a PMMA-mediated process using a dummy (transfer-only) Si pedestal chip. 
This PMMA-based process avoids PDMS and polycarbonate adhesion layer as used previously~\cite{dibos2018ass}. 
We spin a PMMA layer (Microchem A2 950k molecular weight) directly onto a lithographically defined Si pickup pedestal (few microns in height) without baking. 
The Si pedestal is loaded into the flip-chip bonder (SET FC150) and is brought into contact with the released SOI device block and a normal load of about 1\,kg is applied at room temperature to first break the suspended block of Si cavities with high yield at the aforemention break points. 
To “stencil” the PMMA such that it only remains between the device and pedestal, we perform a 5\,min \Otwo\ plasma clean at 150\,W, which removes exposed PMMA on the Si pedestal while leaving PMMA confined under the devices. 
For transfer to the SiC edge, the stamped device block is aligned using lithographic flagposts to register the overhanging Si taper with the SiC chip edge. 
To increase the tolerance of lateral displacement during the stamping process along the pedestal edge, the routing waveguide runs along the SiC surface for 5\um\ before tapering into free space.
We apply 2\,kg normal load and heat both the SiC chip and the Si pedestal to 130\,\degC, near the PMMA reflow temperature. 
This softens the resist and promotes plasma-activated oxide-to-oxide bonding between the Si waveguide surface and native oxide on the SiC pedestal.
A final \Otwo\ plasma stripping step removes residual PMMA.

Typical placement accuracy for the Si nanobeam blocks onto the SiC is $< 2$\,\um\ due the intrinsic accuracy of the flip-chip tool. 
The success rate for the overall device assembly process is critically dependent upon the SiC surface roughness as a rougher surface leads to final step bonding failure. 
However, the device yield for our non-implanted samples with smoother surfaces can be up to 95\,\%. 

\section{Device Design and Characterization}
\label{SIsec:Design}

\begin{figure}[ht]
    \centering
    \includegraphics[width=0.7\linewidth]{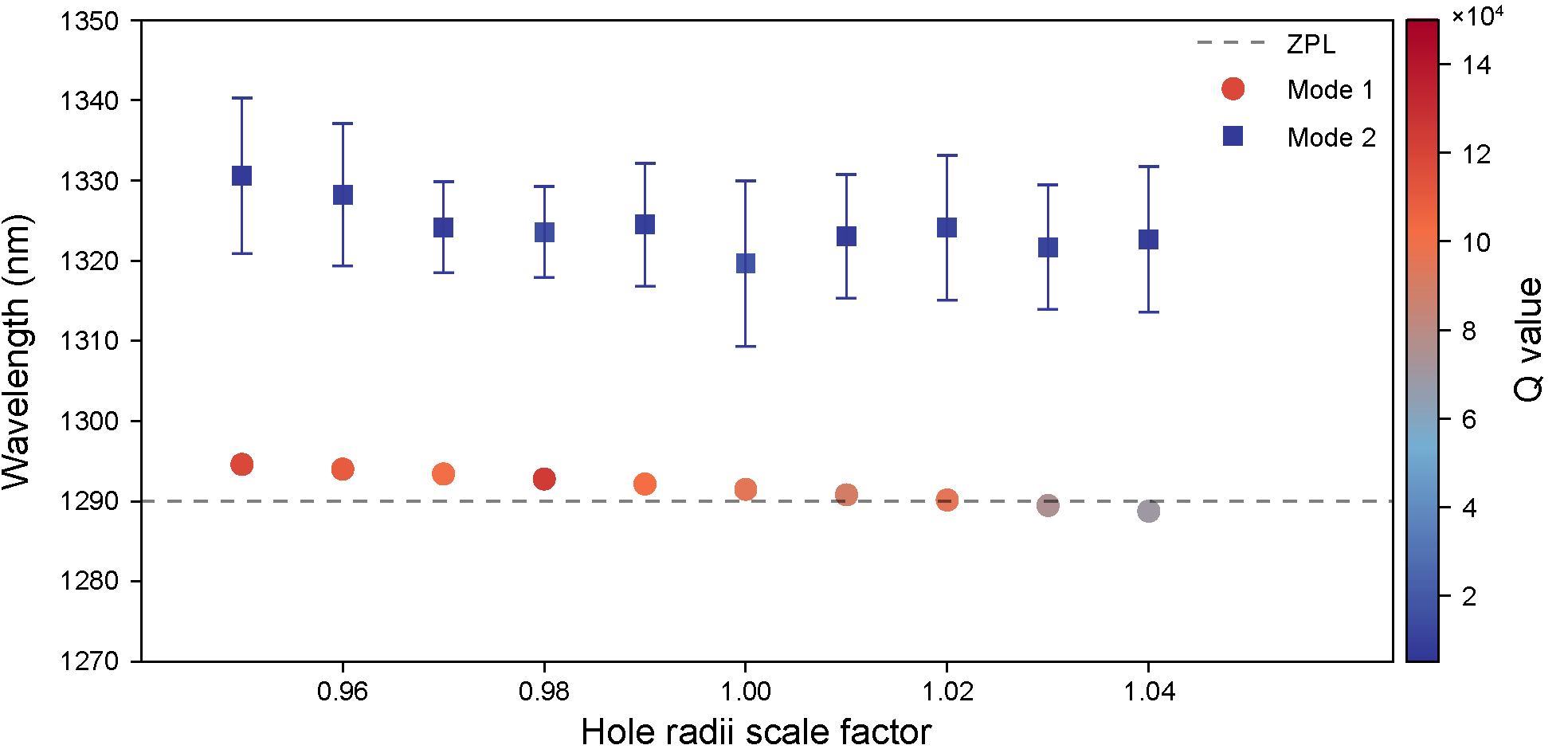}
    \caption{Simulated resonance wavelength and quality factor of our photonic crystal nanocavity design as the size of the elliptical cavity holes is scaled.}
    \label{figS:sim_scale_sweep}
\end{figure}

The starting parameters for our 1D photonic crystal cavity design were chosen using frequency-domain photonic band simulations in MIT Photonic Bands (MPB) to maximize the photonic bandgap. 
A 36- hole cavity defect on a 700\,nm wide waveguide is created by tapering the lattice constant of holes in the waveguides from 221\,nm in the center of the cavity to 235\,nm at the mirrors. 
The tapering of the lattice constant follows a functional form with smooth derivatives to mitigate scattering~\cite{chanThesis2012}. 
The elliptical holes along the cavity are identical in size with a nominal major diameter of 260\,nm and minor diameter of 82\,nm. 
To better understand how experimental variations in these designs affect cavity performance and to further optimize our designs for ongoing work, we performed 3D FDTD simulations using Tidy3D. 
We use an automatic nonuniform grid with minimum 30 steps per wavelength, except in the area surrounding the holes, which has a mesh override of 8\,nm~$\times$~8\,nm~$\times$~20\,nm. 
We simulate an emitter with a TE-polarized point dipole source at the center of the cavity. 
We place a field time monitor at the cavity center and feed the time-domain data to Tidy3D’s build-in ResonanceFinder functionality (an algorithm derived from Ref.~\cite{Mandelshtam1997}) to extract cavity resonances and Q-factors. 
We then place a 3D field monitor across the simulation to extract the field profile at the cavity resonance.

The parameters optimized in MPB will not necessarily result in a cavity exactly at our target resonance and our Si etch rate can vary slightly run-to-run. Therefore, we explored in more detail how slight changes in the elliptical hole size can shift the resonance position and impact Q. 
We uniformly scaled the designed hole major and minor radii by a factor ranging from 0.95 to 1.04 and extracted the first two cavity modes; sweeping across this range shifts the cavity resonance by 5.8\,nm. 
We found that a scale factor of 1.02 results in a resonance at our target wavelength, and the cavity Q peaks at a factor of 0.98 (Figure~\ref{figS:sim_scale_sweep}). 
The error bars for the Q values (most visible for the higher order mode, Mode 2) are generated by sweeping the range of the ResonanceFinder module in Tidy3D and computing a standard deviation of the resonant wavelengths and the average cavity Q from that analysis.

\begin{figure}[ht]
    \centering
    \includegraphics[width=0.9\linewidth]{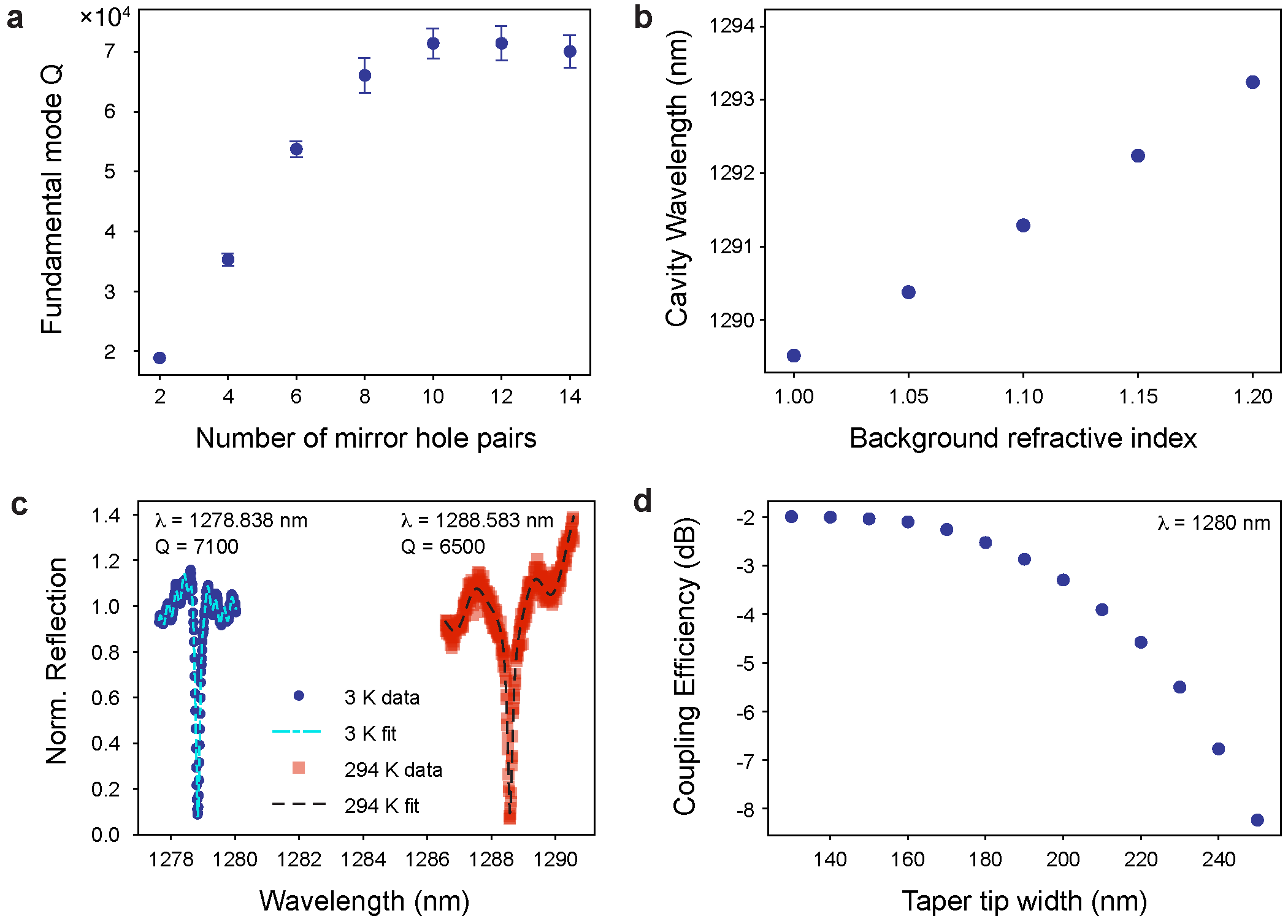}
    \caption{
     \textbf{(a)}\; Simulation of the cavity quality factor as a function of mirror hole number for the particular device dimensions for the cavity investigated in Figure~\ref{fig2:bulk} of the main text.
     \textbf{(b)}\; Simulated shift in cavity resonance wavelength by changing the background refractive index of the simulation.
    \textbf{(c)}\; Example reflection spectra at room (294\,K; blue circles) and cryogenic (3\,K; red squares) temperatures of a cavity stamped on Sample~B. Data are fitted with a Lorentzian on top of an oscillating background due to Fabry-Perot reflections in the experimental fiber circuit.
    \textbf{(b)}\; Simulated waveguide-to-lensed-fiber coupling efficiency at $\lambda$ = 1280 nm as a function of the width of the end of the tapered Si waveguide.
    }
    \label{figS:tuning_taperCoupling}
\end{figure}

Device etching using the \SFsix/\CfourFeight\ etch was challenging, upon inspection via scanning electron microscopy (SEM) the resultant ellipses were distorted: the minor diameter was consistently 92\% wider than the design, as opposed to the major diameter and beam width that were much closer to the target values. We were forced to iterate and use updated hole correction factors to hit the desired GS1-ES1 resonance. The exact cause of this feature distortion for the \SFsix/\CfourFeight\ etch used is currently unclear, but it is likely related to the fluorinated passivation layer created during the etch.

We performed SEM again for the specific nanocavity coupled to the bulk V-doped substrate (used for measurements on Sample~A, Figure~\ref{fig2:bulk} of the main text). We used these extracted parameters in a series of updated Tidy3D simulations to look at the expected Q per mirror hole number for our as-fabricated device geometry (Figure~\ref{figS:tuning_taperCoupling}a). 
Similar to Figure~\ref{figS:sim_scale_sweep} above, the error bars are generated by sweeping the range of the ResonanceFinder module in the Tidy3D. 
There is a saturation of the expected cavity Q for these etched features near $7\times 10^{5}$ for devices with 10 mirror hole pairs. However, for the experimental demonstrations in the main text and elsewhere in the SI, we used only cavities with 4 mirror hole pairs to preserve decent waveguide-to-cavity coupling efficiency due to the aforementioned SiC surface roughness. It should be noted that we experimentally measured cavities with Q-factors up to $3.2 \times 10^{4}$ for smooth bulk V-doped SiC chips with 6 mirror hole pairs, but the cavity resonances were just out of range to be tuned to the V:SiC GS1-ES1 transition.

Critically, we also need to experimentally tune our cavities onto resonance at 3\,K using \Ntwo\ gas condensation. 
We simulated the asymptotic limit of very thick \Ntwo\ gas condensation on a cavity by merely changing the background refractive index of the entire simulation space (Figure~\ref{figS:tuning_taperCoupling}b). 
Nitrogen ice has an approximate refractive index of 1.2~\cite{Yarnall2022}, therefore our simulations suggest a maximum redshift tuning range of approximately 3.8\,nm. 
This range is significantly smaller than the 12\,nm reported for Si-on-\SiOtwo\ devices using \Ntwo\ ~\cite{dibos2022pee} because of the large refractive index of the SiC substrate. 

This gas condensation redshift needs to be balanced with the natural blueshift of experimental resonances upon cool down to 3\,K due to the changing refractive indices of the Si and SiC (Figure~\ref{figS:tuning_taperCoupling}c). 
We can see the same cavity resonance at each of the two temperatures and a net blueshift of roughly 10\,nm for the cooldown process. 
It is important to note that reflection spectra are fit with a Lorentzian to extract the Q and additional terms to account for the sinusoidal modulations caused by reflections in the fiber circuit, similar to Figure~\ref{fig2:bulk} in the main text.

Finally, in FDTD simulations we also swept the end facet width of the inverse taper coupler on the 700\,nm wide Si waveguide to characterize coupling efficiency (Figure~\ref{figS:tuning_taperCoupling}d). 
In these simulations, a Gaussian source is focused at the variable-width tip of a 14\,\um\ linear taper with a 10\,\um\ undercut, and field monitors are placed at the input and output of the waveguide, and a mode monitor is used to find the power coupled into the fundamental TE mode. 
We find that the coupling efficiency at 1280\,nm saturates around -2\,dB ($\sim 63$\,\%) near an input width of 160\,nm. 
We do measure experimental values of 67\,\% for dilute implanted Sample~B, and the 4\,\% underestimate for the simulations might be due to our actual laser-cut lensed optical fiber's geometric parameters (spot size and effective focal length) not being fully captured by our FDTD source.

\section{Experimental Details}

\subsection{Detailed Measurement Configuration}
\label{SIsubsec:config}

\begin{figure}[ht]
    \centering
    \includegraphics[width=0.8\linewidth]{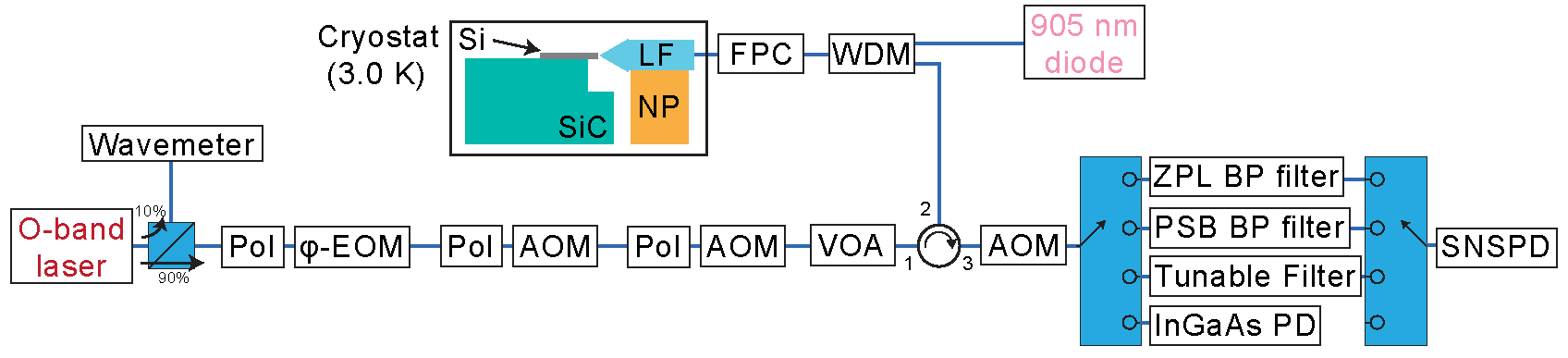}
    \caption{
     A complete diagram of the measurement configuration.
    }
    \label{figS:config}
\end{figure}

In this section we provide a more complete description of the experimental measurement configuration and specific components. 
We use a widely tunable O-band laser (Santec TSL-570) that is monitored with a wavemeter (High Finesse, WS-8). 
This laser is in series with a phase electro-optic modulator (EOM, iXBlue MPZ-LN-10) for sideband modulation and two acousto-optic modulators (AOMs, AA Opto-Electronic MT110-MIR20-Fio-PM0) to produce short pulses of resonant light. 
Each modulator has its own fiber polarizer (POL, Thorlabs ILP1310PM-APC) to preserve high polarization extinction. 
Overall resonant laser power is controlled with a pair of electronic variable optical attenuators (VOA, Thorlabs V1000PA). 
An AR-coated SMF-28-lensed fiber (LF, OZ Optics) is mounted on a 3-axis nanopositioner (NP, Attocube) and couples light to and from the nanocavity inside a closed cycle cryostat (Montana Instruments S100) at 3\,K. 
A wavelength division multiplexer (WDM, Thorlabs custom) is used to inject 905\,nm laser light from a diode (QPhotonics QFLQ-905-200S) with an electronic driver (ALPhANOV CC-S) to generate pulses for V dopant charge state repumping. 
We use an electronic fiber polarization controller (FPC, Phoenix Photonics IEPC-13-1-1-2) to align the polarization to the nanocavity. 
An $8\times8$ fiber optic switch (HUBER+SUHNER Polatis) directs photons to an InGaAs photodiode (PD, Femto) for reflection monitoring or through one of three spectral filter channels depending on the experiment. 
These filter channels consist of 
(i) two ZPL bandpass filters in series each with 1275\,nm center wavelength and 50\,nm wide pass band(Edmund Optics 87-865), 
(ii) two PSB 1300 nm longpass filters (Thorlabs FELH1300) in series, or 
(iii) an electronic tunable narrowband filter (EXFO, XTA-50-O-S-58). 
The free-space filters are mounted in a Thorlabs fiber-to-fiber U-Bench (FBC-1310-APC and FBC-C-FC for the ZPL and PSB filters respectively). 
Each of the filter channels can be routed via the switch to a fiber-coupled superconducting nanowire single-photon detector (SNSPD, Quantum Opus). 
Another identical model AOM in the collection path prevents SNSPD saturation via resonant laser pulses. 
PL and lifetime measurements are performed using a qutools quTAG as a time-tagger.
Single photon time tagging for pulsed autocorrelation measurements is performed with Swabian Instruments Time Tagger Ultra to limit the storage size of the raw datasets. 
A Quantum Machines OPX is used for fast EOM frequency modulation and time tagging for TSHB measurements.

\subsection{Sample Comparison}
\label{SIsubsec:samples}

\begin{table}[ht]
\centering
\small
\renewcommand{\arraystretch}{1.0}
\begin{tabularx}{\linewidth}{@{} Y{1} Y{1} Y{1} @{}}
\toprule
\textbf{Parameter} & \textbf{Sample~A} & \textbf{Sample~B} \\
\midrule
SiC substrate
  & As-grown V-doped, semi-insulating
  & HPSI + V implantation \\
\midrule
V concentration
  & $1.9 \times 10^{17}$ \cmnthree\ ($\sim4$\,ppm), uniform
  & $1.9 \times 10^{11}$ \cmntwo\ fluence, 150\,nm mean depth \\
\midrule
Post-growth processing
  & Pedestal etch only
  & Pedestal etch, 275\,keV V implant at 500\,\degC, 1400\,\degC\ Ar anneal (30\,min, C-cap), 750\,\degC\ air anneal (60 min) \\
\midrule
Surface roughness (post-processing)
  & As received (sub-nm RMS)
  & $\sim1$\,nm RMS \\
\midrule
Representative Q
  & $1.2 \times 10^{4}$
  & $7.9 \times 10^{3}$ \\
\midrule
Cavity-waveguide coupling ($\eta_\mathrm{CW}$)
  & 0.27
  & 0.35 \\
\midrule
Waveguide-fiber coupling
  & 0.18
  & 0.67 \\
\midrule
Measurement type
  & Ensemble V
  & Single V \\
\bottomrule
\end{tabularx}
\caption{Key parameters for Samples A and B.}
\label{tab:samples}
\end{table}

\section{Purcell-Enhanced Ensemble Characterization}
\label{SIsec:ensemble}

\subsection{Lifetime Parameters as a Function of Cavity Detuning}
\label{SIsec:ensemble-lifetimes}

\begin{figure}[ht]
    \centering
    \includegraphics[width=0.78\linewidth]{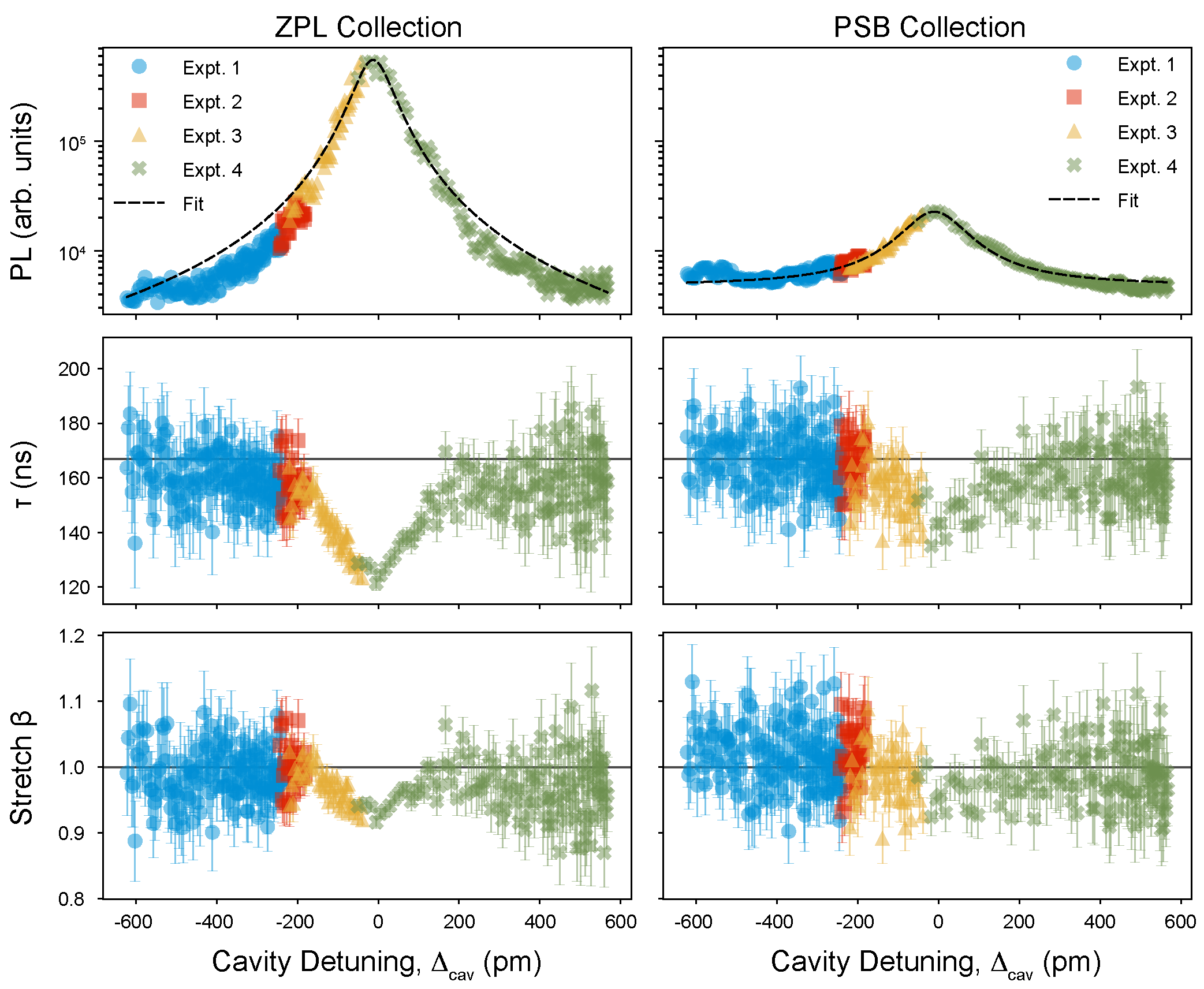}
    \caption{Summary of stretched-exponential-fitted time-resolved PLE traces as a function of cavity detuning. Each dataset is represented by a different color and marker. Data were collected at 3\,K, under resonant excitation at 130\,nW for 50\,ns and collection window of 800\,ns. Left: ZPL-filtered data; right: PSB-filtered data (top)\; Total accumulated photoluminescence. (middle)\;Characteristic time $\tau$ of a stretched exponential fit. (bottom)\;Stretch parameter $\beta$ from the same fit.}
    \label{figS:lifetime-summary}
\end{figure}

As we show in Figure~\ref{fig2:bulk} and describe in the main text, we quantify the cavity-mediated emission enhancement by slowly sweeping the cavity resonance across the GS1-ES1 transition while monitoring the collected photon counts under resonant excitation. 
We alternate between the ZPL and PSB filters via a double-loop on the $8 \times 8$ optical switch to independently track the cavity’s effect on each emission channel.
When the cavity is far-detuned from the GS1-ES1 transition, the excitation pulse is predominantly reflected and only addresses bulk emitters coupled directly underneath the waveguide, as the emitters underneath the cavity region are not efficiently excited because the laser wavelength is in the photonic stopband.
Under these off-resonance conditions, the detected ZPL and PSB count rates are approximately equal; an apparent discrepancy with the reported Debye-Waller factor of 0.25 for bulk V:SiC~\cite{wolfowicz2020vsq}, which may be attributed to wavelength-dependent differences in waveguide-to-fiber out-coupling efficiency, optical circuit transmission, and SNSPD detection efficiency between the two spectral windows.
When the cavity is tuned onto resonance, the ZPL emission increases by up to $121\times$ relative to the far-detuned background, with an enhancement linewidth matching the cavity linewidth (100\,pm). 
In contrast, the PSB emission shows only a $4.1\times$ enhancement with a broader linewidth of 178\,pm. 
Concurrently, the radiative lifetime measured on resonance decreases from the bulk value of 167\,ns~\cite{wolfowicz2020vsq} to 121\,ns for ZPL-filtered emission (stretched-exponential exponent ($\beta=0.92$; $I\propto e^{\left(-t/\tau\right)^\beta}$) and to 140\,ns for PSB-filtered emission ($\beta=0.95$).
The large ZPL count-rate enhancement therefore reflects two combined effects: (i) an increased number of resonantly excited emitters that couple to the cavity mode, and (ii) a Purcell-enhanced spontaneous emission rate into the ZPL channel. 
The modest PSB enhancement, by contrast, arises primarily from the larger excited population, partially offset by cavity-induced suppression of off-resonant emission channels and by non-optimized emitter-cavity and cavity-waveguide coupling at PSB wavelengths. 
We do not fully understand the modest difference in the fit stretched exponential time constants nor effective cavity-modified linewidths between the ZPL and PSB filtered signals and both are a subject of further study. 
Because these population increase and rate enhancement contributions cannot be fully disentangled in an ensemble measurement, we estimate an upper bound on the Purcell factor by taking the ratio of the ZPL to PSB count-rate enhancements (and correcting for background counts), $F_P \leq 120 / 3.1 \simeq 38$, under the assumption that both channels experience the same increase in excited population.

\subsection{Emission Spectrum as a Function of Cavity Detuning}
\label{SIsubsec:ensemble-coll-dep}

\begin{figure}[ht]
    \centering
    \includegraphics[width=0.7\linewidth]{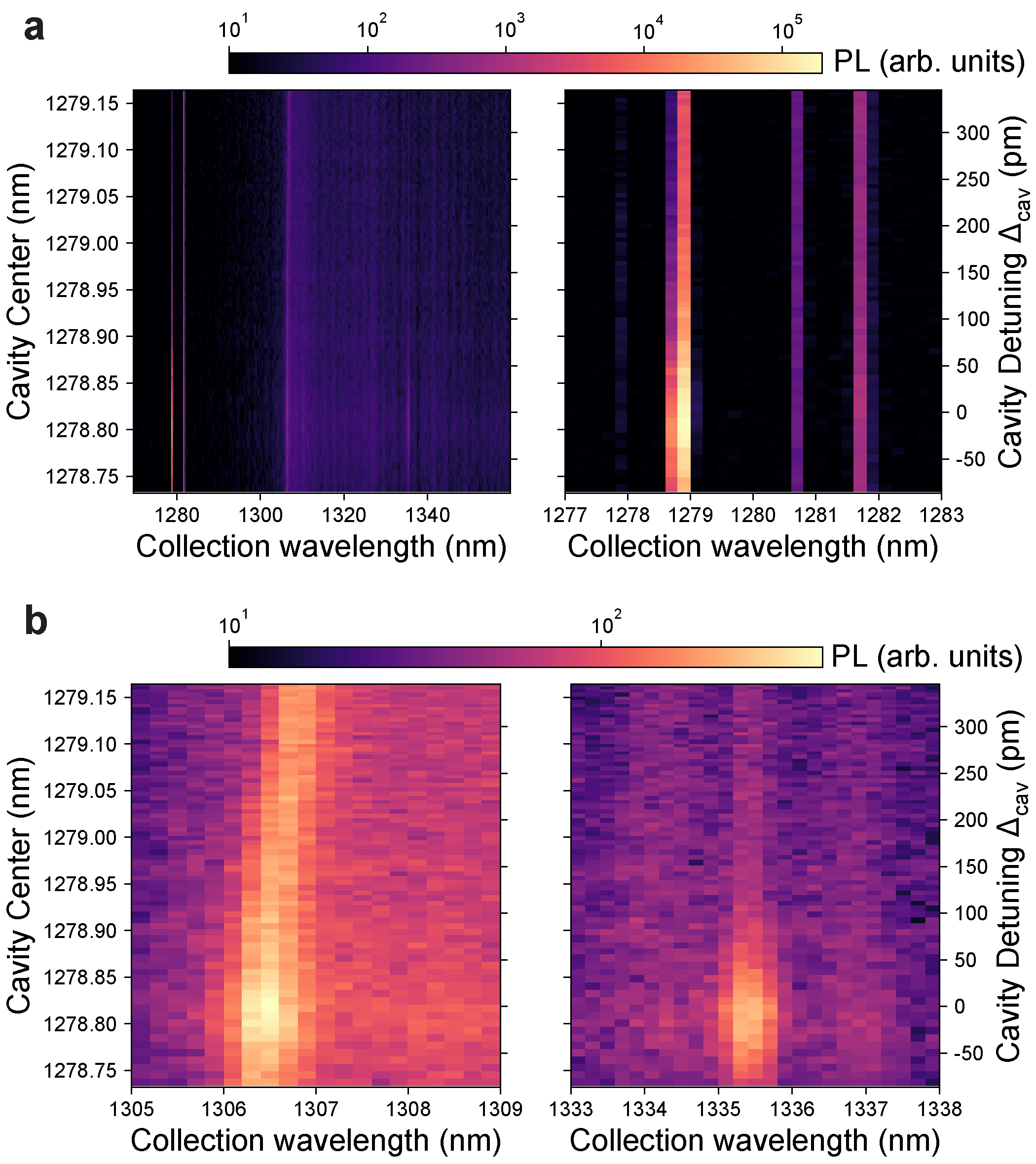}
    \caption{
    Total PLE emission as a function of collection wavelength (x-axis) and cavity detuning (y-axis), under resonant excitation at 2.5\,\uW\ for 50\,ns and collection window of 800\,ns at 3\,K. 
    (a)\; (Left) Spectra in the full collection wavelength range and (Right) focused on the ZPL range $1277 - 1283$\,nm, sharing the same colorscale.
    (b)\; (Left) Spectra focused on the feature near $1305-1309$\,nm which is attributed to a higher order cavity mode. (Right) Spectra focused on the \thh\ site $1333 - 1338$\,nm.
    }
    \label{figS:drift-col-ple}
\end{figure}

The PL emission spectra depicted in Figure~\ref{fig2:bulk}e were collected in a series of cavity sweep measurements with three particular slices of the laser-cavity detuning ($\Delta_{cav}$). 
Figure~\ref{figS:drift-col-ple} shows the full two-dimensional map of the PLE emission for the full range of $\Delta_{cav}$ under resonant GS1-ES1 excitation. 
These are plotted at different filter collection wavelengths (x-axis) versus varying cavity position (left y-axis) or $\Delta_{cav}$ (right y-axis). 
We observe four prominent features: (i) four ZPL transitions (GS1-ES1, GS1-ES2, GS2-ES1, and GS2-ES2), (ii) a broad PSB spectrum for wavelengths longer than 1300\,nm, (iii) a sharp feature near the beginning of the PSB near 1306.5\,nm and (iv) another one sitting atop the PSB at 1335.5\,nm.
The feature near 1306.5\,nm has been reported previously to be a vibronic mode~\cite{spindlberger2019opv}, and this peak is further enhanced in our measurements by a relatively broad higher-order cavity mode (Mode 2 in Figure~\ref{figS:sim_scale_sweep}) that spectrally overlaps with it; evident by the spectral drift observed in Figure~\ref{figS:drift-col-ple}b. 
The feature near 1335.5\,nm is emission from the \thh\ site, which appears as the cavity tunes onto the GS1-ES1 resonance, suggesting that the enhanced intracavity field enables non-resonant excitation of this secondary crystallographic site.

In the main text, we outline that from these measurements we compute the percentage of detected photons originating from the GS1-ES1 ZPL. When the cavity is far-detuned, the GS1-ES1 line accounts for 90\,\% of the total ZPL emission; on resonance, the cavity funnels nearly all ZPL emission into this single line (99.4\,\%). Both of these percentages are calculated from the on-resonance spectrum in Figure~\ref{fig2:bulk}e in the main text, integrating the GS1-ES1 line versus the sum of all four ZPL transitions while subtracting the background contribution.

\subsection{Transient Spectral Hole Burning as a Function of Resonant Power}
\label{SIsec:TSHB}

\begin{figure}[ht]
    \centering
    \includegraphics[width=0.8\linewidth]{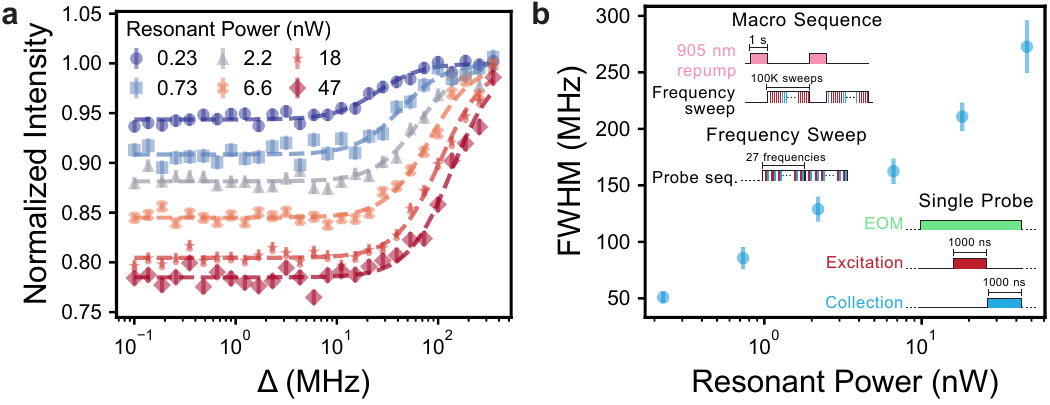}
    \caption{$T=3$\,K, under resonant excitation: (a)\;Transient spectral hole burning spectra of as-grown V ensemble resonant with the nanocavity, showing the normalized PL intensity as a function of the EOM carrier-sideband detuning ($\Delta$) for different resonant excitation powers.
    (b)\; Full-width at half-maximum fit results from (a) as a function of resonant excitation power. Inset schematics show the pulse sequences in the TSHB experiment: (i) a single laser pulse sequence for the EOM drive, resonant excitation, and collection, (ii) A single frequency sweep where the frequencies are Fisher-Yates-shuffled, and (iii) the associated larger experimental sequence whereby the charge repump pulses are used before hundereds of thousands of successive frequency sweeps.}
    \label{figS:tshb-vs-power}
\end{figure}

As discussed in the main text, we perform transient spectral hole burning measurements as a proxy for the homogeneous linewidth of individual emitters when part of an ensemble. To do so, we insert optional phase-electro-optic modulator ($\phi$-EOM dashed box within Figure~\ref{fig2:bulk}a) into the excitation path to generate sidebands at variable frequencies ($\Delta$) around the carrier. 
After calibrating the power output of the RF so that the sidebands are equal to the carrier amplitude for each frequency, we perform single-probe measurements by exciting the ensemble for 1000\,ns and collecting the subsequent PL for 1000\,ns. 
We sweep the sideband detuning across $N$ points (from 0.1 to 400\,MHz) within a window of 60\,\us, randomizing the frequency sequence in each cycle to eliminate intensity fluctuation artifacts, and repeat this sweep $\sim2.5\times10^6/N$ times between 905\,nm charge-repump pulses. 
We fit the data with a negative Lorentzian profile under a constant background and normalize both fit and data to the fitted background value (Figure~\ref{figS:tshb-vs-power}a).

At high excitation powers, the hole-burning contrast can reach a theoretical maximum contrast of $\sqrt{3}/3\approx 0.577$, but the linewidth is significantly broadened by power-induced dephasing. 
By reducing the excitation power, we mitigate this broadening at the expense of signal contrast, which was further challenged by a first-generation waveguide taper design that limited fiber-coupling efficiency to 18\,\% for Sample~A. 
The trend of FWHM versus resonant power (Figure~\ref{figS:tshb-vs-power}b) does not show that 50\,MHz is the lower-most linewidth achievable, and may be improved with bulk V devices with higher cavity-to-waveguide and waveguide-to-fiber collection efficiencies.  
Regardless of the resonant pump power, the TSHB linewidth is always well below the linewidth of the cavity ($\sim 19$\,GHz), which places the emitters firmly in the bad cavity limit, such that the Purcell factor is not limited by the optical linewidth of the emitters. 

\section{Charge Dynamics}
\label{SIsec:charge}

\subsection{Experiments Without Charge Repumping}
\label{SIsubsec:no_repump}

\begin{figure}[ht]
    \centering
    \includegraphics[width=0.8\linewidth]{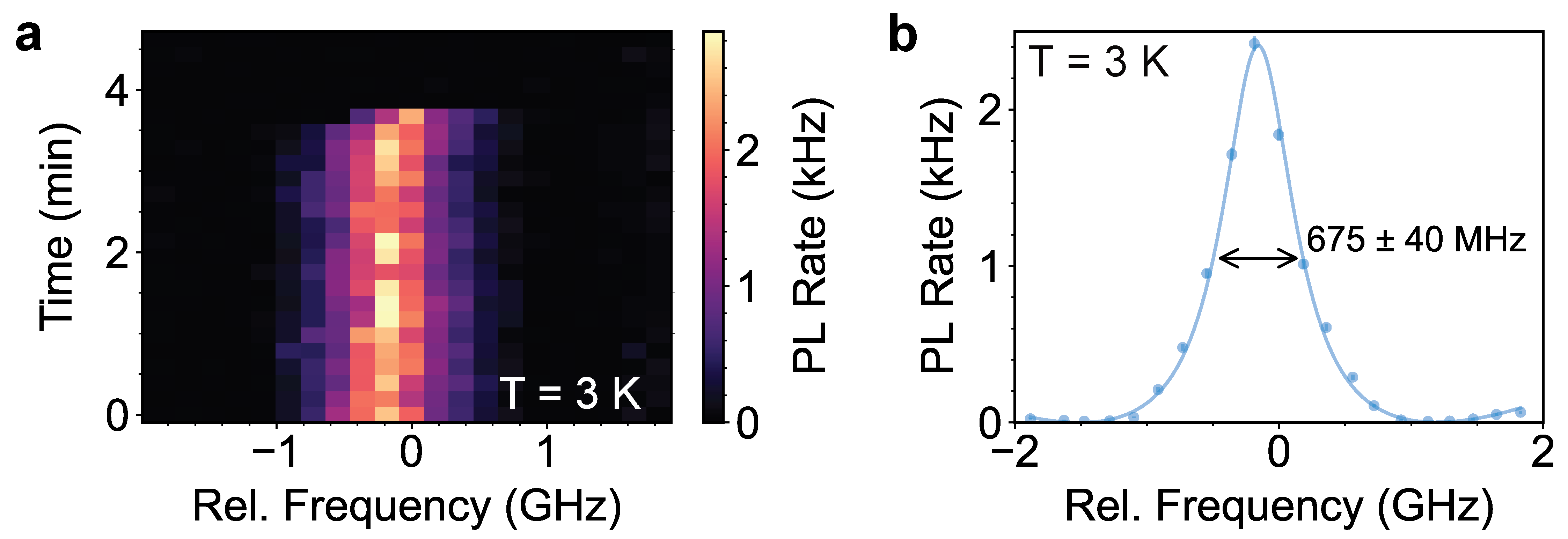}
    \caption{
    \textbf{(a)} (Left axis) PLE spectroscopy over time of single emitter (SE1) resonant with the cavity with intensity displayed in units of photons detected per second of collection time without the use of the 905\,nm repump laser.
    For each wavelength, we probe 250k times at 240\,pW for 50\,ns, collect for 250\,ns, with a total cycle time of 440\,ns. 
    \textbf{(b)} An average of the intensity for the entire measurement duration in (a) with a Lorentzian fit.    
    }
    \label{figS:no_repump}
\end{figure}

We initially performed similar resonant scanning on SE1 to that shown in Figure~\ref{fig4:single} of the main text, but without the use a repump laser (Figure~\ref{figS:no_repump}). 
Without the repump, the emitter linewidth is slightly narrower but after some period of probing ionizes and does not return to the \Vo\ charge state (Figure~\ref{figS:no_repump}). 

\subsection{Optimizing Charge Repumping Conditions}
\label{SIsubsec:repump}

\begin{figure}[ht]
    \centering
    \includegraphics[width=0.9\linewidth]{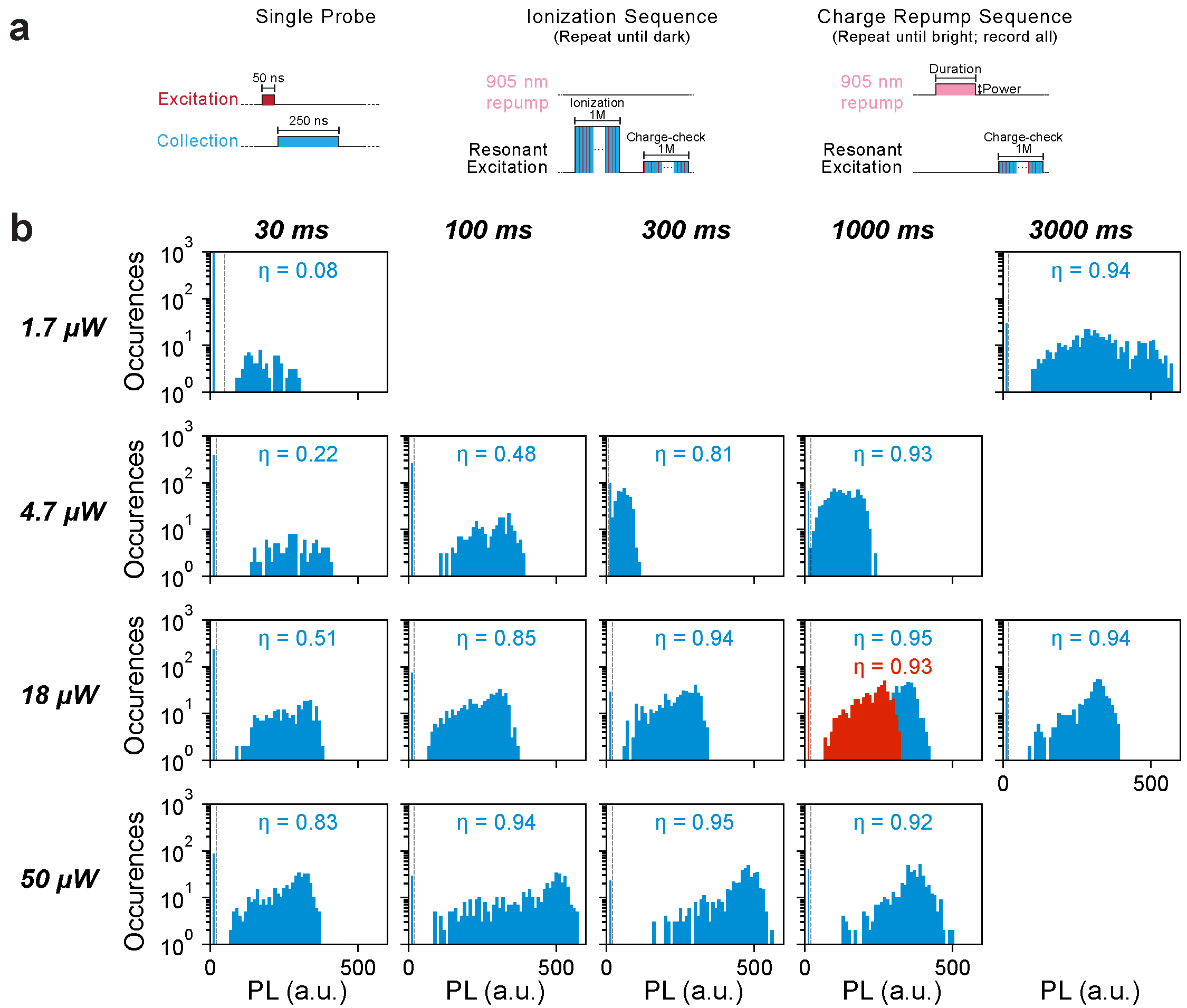}
    \caption{
    \textbf{(a)} Pulse sequences for the repump efficiency measurement.
    \textbf{(b)} Measurement of 905\,nm repump efficiency on single emitter SE1. Histograms of number of photons recorded during 1 million probe pulses following a 905\,nm repump pulse is applied to SE1 in its dark charge state. The power and duration of the repump pulse are displayed to the left and top of each histogram respectively. 
    The resonant pulse is set to 240\,pW for a Charge-check and 52\,nW for Ionization. The total time of a single probe sequnce is 440\,ns. 
    The discrimination threshold between the light and dark charge states is shown as a vertical dashed line in each histogram and the resulting measured repump efficiency is shown as $\eta$. The red histogram at 18\,\uW\ and 1000\,ms repump pulse time was taken at the same repump conditions as the blue histogram but at a slightly different laser-cavity detuning. All data were collected at $T=3$\,K.
    }
    \label{figS:repump-eff-table}
\end{figure}

The ionization results spurred us to develop a deterministic ionization-recovery protocol compatible with our hybrid Si-on-SiC device geometry.
We employed three pulse sub-sequences (\textit{Single Probe, Ionization, and Charge Repump}) shown in Figure~\ref{figS:repump-eff-table}a  in the our repump experiments.
The experiment starts by ensuring the color center is in the dark state by pumping with strong resonant excitation to replicate the effect shown in Figure~\ref{figS:no_repump}a but without laser frequency scanning.
We then confirm the defect is in the dark state via a low-power charge-state check consisting of one million \textit{Single Probe} cycles (\textit{Ionization Sequence}).
The total counts observed during the Charge-check are used to discriminate between the dark and bright charge states. 
Once the defect is observed in the dark charge state, the \textit{Charge Repump Sequence} is applied until the bright state is restored. 
The \textit{Charge Repump Sequence} consists of a 905\,nm laser pulse at a specific power and duration followed by the same Charge-check step from the \textit{Ionization Sequence}. 
If the defect is observed to be in the dark state after the \textit{Charge Repump Sequence}, the 
\textit{Charge Repump Sequence} is repeated, but if the bright state is observed, the experiment switches back to the \textit{Ionization sequence}.
The brightness results from each \textit{Charge Repump Sequence} are recorded and the experiment continues until 500 total \textit{Charge Repump Sequences} have been completed.
The results at various 905\,nm repump power and pulse durations are shown in Figure~\ref{figS:repump-eff-table}b.

We calculate the repump efficiency for each combination of laser power and pulse duration as the fraction of post-repump Charge-check pulses that exceed a fixed count threshold and the results are shown in Figure~\ref{fig3:charge}c of the main text.
The repump efficiency $\eta_\mathrm{rp}$ is well described by a single-exponential fit to the total delivered repump energy (power $\times$ duration), following $\eta_\mathrm{rp} = \alpha (1-e^{-\gamma_\mathrm{rp} I\tau})$ where $\alpha$ is the saturated bright charge state occupation, $\gamma_\mathrm{rp}$ is the energy repump coefficient, $I$ is the delivered optical power, and $\tau$ is the repump pulse duration.
The maximum observed repump efficiency saturates at 95\,\%, with the residual 5\,\% potentially arising from back-conversion to the dark state during the repump pulse itself.

The fact that the fitting model only depends on the total repump energy and not on the instantaneous repump intensity implies that the repump process likely only requires a single repump photon (though a two photon process is still possible if the two photons drive different processes and only one process is rate limiting).
While this does not definitively establish which charge state is the dark state, it does imply it may be \Vn.
The \Vn/\Vo charge transition level has been measured to occur at an energy of $E_C - 1.11$\,eV~\cite{mitchel2007vda}, which means that a single 905\,nm photon (1.37\,eV) is sufficient to remove an electron from a defect in the \Vn\ state into the conduction band -- transforming it into the \Vo\ state in the process. 
In contrast, the \Vo/\Vp\ charge transition level is closer to the center of the band gap $E_C - 1.57$\,eV ($E_V + 1.69$\,eV)~\cite{mitchel2007vda} so it would not be possible for a single 905\,nm photon to directly ionize the defect via either the valence or conduction band.
This implies that if the dark state is the \Vp\ state, then the repump process would most likely require the involvement of another background defect.
For example, repumping might occur by photogeneration of an electron from a shallow donor into the conduction band followed by charge capture by the positivity charged \Vp\ defect.
If this were the case, the repump efficiency from 905\,nm light might vary significantly between defects with different neighborhoods or in differently doped host materials.

\subsection{Optimizing Resonant Probing Conditions}
\label{SIsubsec:resonant_pumping}

\begin{figure}[ht]
    \centering
    \includegraphics[width=0.7\linewidth]{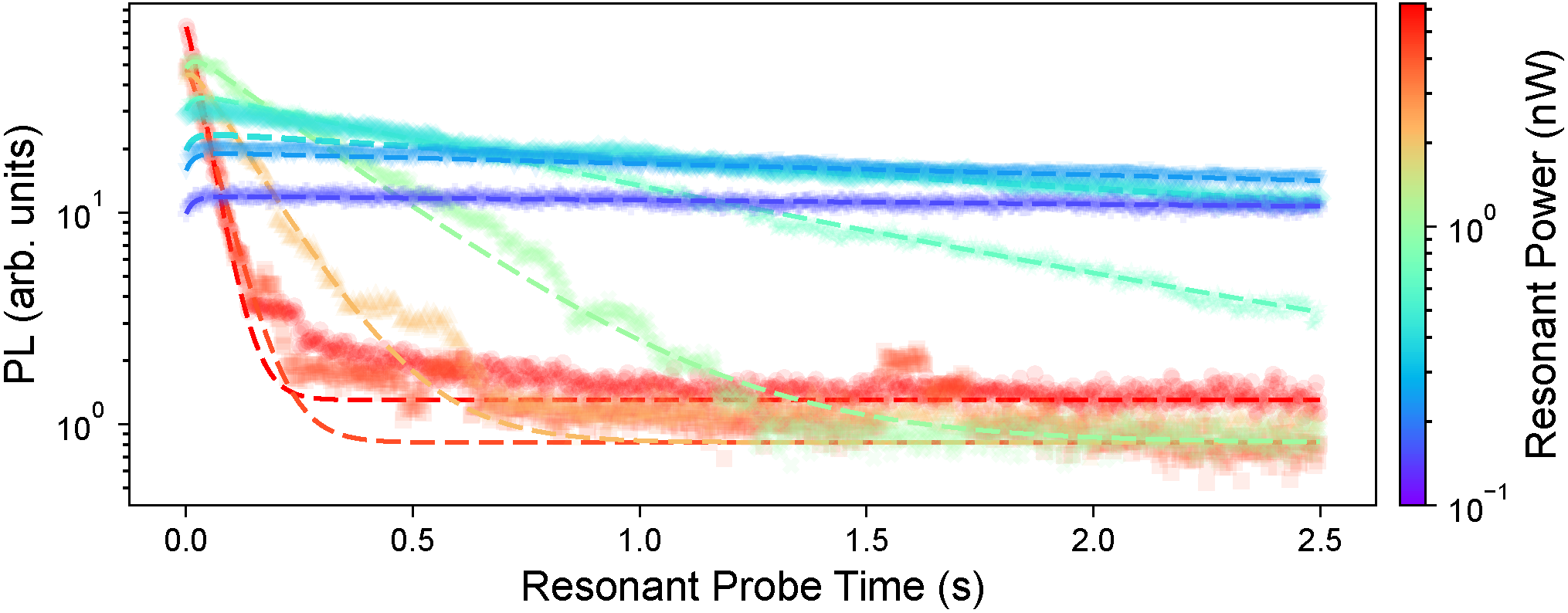}
    \caption{ 
    Measurement of ionization rate for different resonant illumination intensities. Each colored curve shows the average brightness of SE1 as a function of exposure time to resonant excitation. 
    For each resonant power, in a given Charge-Repump Sequence we probe 50 million times for 50\,ns, collect for 250\,ns, with a total cycle time of 440\,ns. 
    The repump power was 4.7\,\uW\ and duration was 1\,s
    The color of each curve corresponds to the intensity of the resonant light according to the colorbar. Each measurement was preceded by a charge repump pulse. The dashed lines are an exponential fit to the decay curves which are used to extract the ionization rate shown in main text Figure~\ref{fig3:charge}. The fit also includes a slight recovery at short times to account for a small wavelength shift of the transition following a repump pulse which is shown in more detail in Figure~\ref{figS:repump-wvl-shift}a.
    }
    \label{figS:ion-power-dep-raw}
\end{figure}

\begin{figure}[ht]
    \centering
    \includegraphics[width=1.0\linewidth]{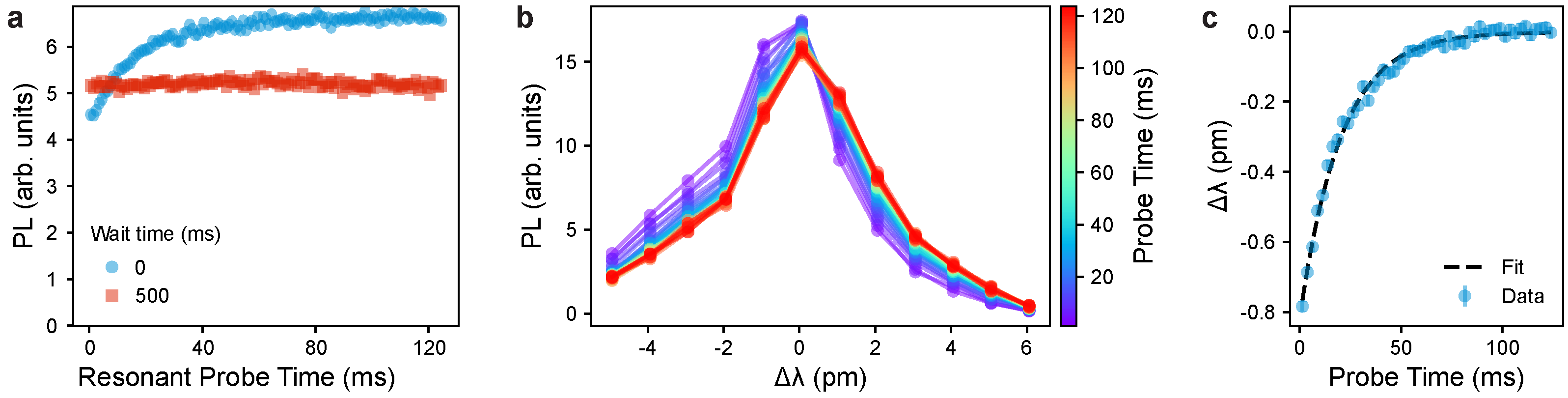}
    \caption{ 
    For each of the following, we probe 2.5 million times at 240\,pW for 50\,ns, collect for 250\,ns, having a total cycle time of 440\,ns. The repump power was 4.7\,\uW\ and duration was 1\,s. Measurements were taken at $T=3$\,K.
    \textbf{(a)} Comparison between probe-time-response of SE1 emission with and without a wait-time between the end of the repump pulse and the start of the probing sequence.
    \textbf{(b)} Emission spectrum of SE1 for different time delays following a repump pulse. The peak emission wavelength is observed to slightly redshift with increasing time following the repump pulse. \textbf{(c)} Detuning of the center of the emission spectrum (calculated from the weighted mean of each spectrum at a give probe time slice) from the steady state value as a function of probe time following a repump pulse. The dashed line is an exponential fit to the data with a recovery timescale of 20.7\,ms of probe exposure with a 11.4\,\% duty cycle.
    }
    \label{figS:repump-wvl-shift}
\end{figure}

Having established a reliable charge-recovery protocol, we carefully characterize the ionization dynamics under resonant excitation to identify the optimal probe conditions that will balance V dopant brightness with the rate of ionization. 
At a range of resonant excitation powers, we perform a sequence of repump-probe measurements and observe two distinct timescales in the photon count trajectory as a function of probe pulse number (Figure~\ref{figS:ion-power-dep-raw}).  
On longer timescales, an exponential decay reflects the expected photoionization of the \Vo\ center to the dark state, with a rate that increases with excitation power. 
However, on shorter timescales, we observe an unexpected initial exponential recovery following the repump pulse. 
This recovery seems to take place in the absence of active probing, as we do not observe it if a delay of 500\,ms is inserted between the repump and probe pulses (Figure~\ref{figS:repump-wvl-shift}a).
By performing an excitation wavelength sweep while measuring the emitted PL as a function of resonant probe time, we spectrally resolve the recovery process (Figure~\ref{figS:repump-wvl-shift}b).
We observe a behavior that can be described as a resonance distribution with a fixed FWHM and whose center exponentially relaxes to a given steady-state value with an amplitude $\Delta\lambda_0$ and characteristic time $\tau_r$.
We find that the mean GS1-ES1 transition center is initially detuned by $\Delta\lambda_0=-0.8$\,pm ($\sim 140$\,MHz) and shifts back towards its steady state value during the recovery (Figure~\ref{figS:repump-wvl-shift}c).
As a function of probe time -- with probes starting immediately after the end of the repump pulse -- this recovery has a characteristic time scale of $\tau_r=20.7$\,ms (Table~\ref{tab:wvl-shift}).
The probe has an 11.4\,\% duty cycle so the real time recovery timescale is 182\,ms.
The timescale of this recovery also appears to be largely independent of the repump power, as shown in Table~\ref{tab:wvl-shift}.
The apparent increase in brightness in Figure~\ref{figS:ion-power-dep-raw} is therefore an artifact of the defect shifting back onto resonance with the excitation laser.
We attribute this wavelength shift and recovery to disturbance and subsequent relaxation of the local charge environment toward a steady state following the repump pulse.
Returning to the long timescale of Figure~\ref{figS:ion-power-dep-raw}, we are now able to use fixed fit parameters for the short timescale behavior and attain reliable fits for the ionization decay at each resonant power.
The ionization decay follows a power-law dependence with exponent $1.73\pm0.08$ as shown in Figure~\ref{fig3:charge} in the main text.

\begin{table}[ht]
\centering
\small
\begin{tabular}{c|c|c|c}

Repump Power (\uW) & Recovery (ms)   & $\Delta\lambda$ (MHz) & FWHM (MHz) \\ \hline
1.7                & $20.7 \pm  0.7$ & $146 \pm 2$           & 872        \\ \hline
4.7                 & $20.3 \pm 0.6$  & $140 \pm 2$           & 911        \\ \hline
4.7                & $22.5 \pm 1.0$  & $141 \pm 3$           & 737        \\ \hline
18                 & $20.4 \pm 0.8$  & $140 \pm 2$           & 798        \\ \hline \hline
Mean               & $21.0 \pm 0.4$  & $142 \pm 1$           & 829        \\ 
\end{tabular}
\caption{Extracted wavelength recovery parameters from fits such as the one in Figure~\ref{figS:repump-wvl-shift}b for different repump pulse intensities. All repump pulses have a duration of 1\,s. We observe little to no dependence on repump power of the wavelength shift and recovery timescale observed after a repump pulse.}
\label{tab:wvl-shift}
\end{table}

\section{Quantum Efficiency and Photon Budget}
\label{SIsec:QE}

\begin{table}[ht]
\centering
\small
\renewcommand{\arraystretch}{1.0}
\begin{tabularx}{\linewidth}{@{} Y{1} c Y{1} @{}}
\toprule
\textbf{Coupling / Loss Mechanism} & \textbf{Efficiency} & \textbf{Comments} \\
\midrule
Emission probability into GS1-ES1 ZPL ($P_\mathrm{GS1-ES1}$)
  & $\displaystyle\sim\frac{(1+F_P)\,\xi D\eta_\mathrm{QE}}{1 + F_P\,\xi D\eta_\mathrm{QE}}$
  & Detuning-dependent \\
\midrule
Photon coupling from defect to cavity ($\eta_\mathrm{DC}$)
  & $\displaystyle\sim \frac{F_P}{F_P+1}$
  & Detuning-dependent; may be estimated from lifetime change as described below \\
\midrule
Cavity to waveguide coupling ($\eta_\mathrm{CW}$)
  & 0.35
  & Calculated from visibility of cavity reflection spectrum \\
\midrule
Waveguide to lensed fiber coupling       & 0.67   & Measured via reflection    \\
Lensed fiber incl.\ cryostat feedthrough & 0.88   & Measured via transmission  \\
Fiber polarization controller            & 0.8125 & Measured via transmission  \\
Wavelength Division Multiplexer          & 0.792  & Measured via transmission  \\
Optical Circulator                       & 0.80   & Measured via transmission  \\
Collection Path\textsuperscript{*}       & 0.1875 & Measured via transmission  \\
SNSPD detection efficiency               & 0.7    & Not known precisely        \\
\midrule
Total (on resonance)
  & $0.0108 \times P_\mathrm{GS1-ES1} \times \eta_\mathrm{DC}$
  & \\
\bottomrule
\end{tabularx}
\caption{Compilation of coupling efficiencies and loss mechanisms involved in our experimental setup. $F_P$ is the Purcell factor, $\xi$ is the unenhanced fraction of ZPL emission through the GS1-ES1 transition, $D$ is the unenhanced Debeye-Waller factor, and $\eta_\mathrm{QE}$ is the unenhanced radiative quantum efficiency.
Our measurements of Sample~A found $\xi \approx 0.9$ when the cavity is off resonance (Figure~\ref{fig2:bulk}e). Past experiments have estimated $D\approx0.25$ and $\eta_\mathrm{QE}$ within 0.02--0.2 for the \thk\ site~\cite{wolfowicz2020vsq}.
The collection path includes: a fiber AOM, an optical switch, a fiber UBench containing $2\times1275$\,nm bandpass filters, and several meters of fiber to deliver light to the SNSPD}
\label{tab:efficiencies}
\end{table}

\subsection{Estimating detection efficiency of ZPL photons}

An important metric for a defect based spin-photon interface or single photon emitter is the rate at which indistinguishable photons can be extracted from the system.
Part of this is a result of the defect's radiative lifetime, but the fraction of photons emitted in the GS1-ES1 ZPL transition and the efficiency with which those photons can be collected into fiber are also important considerations.
Here, we focus on the probability that a GS1-ES1 photon will couple into fiber and furthermore, once coupled into fiber, the probability that it is detected in our particular measurement setup.
This discussion will use values from the cavity containing SE1, which is in Sample~B.

The efficiency of coupling a photon emitted from a cavity-coupled defect into fiber is primarily set by three factors.
First, is the probability that the photon is emitted into the cavity mode ($\eta_\mathrm{DC}$).
This depends strongly on the achieved Purcell enhancement and goes roughly as:
\begin{equation}
    \eta_\mathrm{DC} \propto \frac{F_P}{F_P+1}.
    \label{eq:defect-cavity}
\end{equation}
Estimates of the Purcell factor ($F_P$) will be discussed below.
The second factor is the cavity to waveguide coupling ($\eta_\mathrm{CW}$), which is controlled by the competition between the cavity's internal loss channels such as absorption and scattering, and the emission rate into the waveguide.
This can be measured through the visibility of the resonance in a reflection spectrum, where for under-coupled cavities:
\begin{equation}
    \eta_\mathrm{CW} = \frac{1}{2}(1-\sqrt{1-V}).
\end{equation}
In our case, we find higher than expected scattering losses -- likely due to surface roughness from annealing -- that result in $V = 0.91$ and therefore $\eta_\mathrm{CW} = 0.35$.
An emitted photon could also to couple into the waveguide mode directly, but for Purcell factors above $\sim 10$ we expect this to be a largely negligible contribution.
The third factor is the waveguide to fiber coupling.
We measure this directly via the reflected power from the device when the laser wavelength is within the photon stop band, but not resonant with the cavity.
We observe a single direction waveguide to fiber coupling efficiency of 0.67 which is in good agreement with the simulated value (see Figure~\ref{figS:tuning_taperCoupling}).

Existing literature and our measurements do not provide enough information to calculate the Purcell factor achieved for SE1 precisely, so we will instead compare several estimates to determine a likely value of $F_P$ for our calculation of the defect-to-fiber coupling efficiency of GS1-ES1 photons.
We will first use the reduction in lifetime as a result of the cavity enhancement and literature values for the non-decay rates for bulk emitters to obtain a range of reasonable Purcell factors.
The total radiative rate of a cavity enhanced emitter is
\begin{equation}
    \Gamma_\mathrm{tot}\left(F_P\right) \simeq \Gamma_\mathrm{GS1-ES1}F_P + \Gamma_\mathrm{ZPL} +  \Gamma_\mathrm{PSB} + \Gamma_\mathrm{NR},
\end{equation}
where $\Gamma_\mathrm{GS1-ES1}$, $\Gamma_\mathrm{ZPL}$, $\Gamma_\mathrm{PSB}$, and $\Gamma_\mathrm{NR}$ are the GS1-ES1 line, total zero phonon line, phonon sideband, and non-radiative decay rates for a bulk emitter respectively and $\Gamma_\mathrm{tot}\left(0\right)$ recovers the unenhanced decay rate.
The four decay rates are related via $\xi$, the ratio of GS1-ES1 emission to all ZPL emission, $D$, the Debeye-Waller factor, and $\eta_\mathrm{QE}$, the radiative quantum efficiency as:
\begin{align}
    \xi &= \frac{\Gamma_\mathrm{GS1-ES1}}{\Gamma_\mathrm{ZPL}}\\
    D &= \frac{\Gamma_\mathrm{ZPL}}{\Gamma_\mathrm{ZPL} + \Gamma_\mathrm{PSB}}\\
    \eta_\mathrm{QE} &= \frac{\Gamma_\mathrm{ZPL} + \Gamma_\mathrm{PSB}}{\Gamma_\mathrm{ZPL} + \Gamma_\mathrm{PSB} + \Gamma_\mathrm{NR}}
\end{align}
Given that we know the radiative lifetime ($\tau_\mathrm{tot}\left(F_P\right) = 1/\Gamma_\mathrm{tot}\left(F_P\right)$ for both bulk emitters (167\,ns) and SE1 when on resonance with the cavity (109 ns) we can solve the above equations for the value of $F_P$ achieved and find:
\begin{equation}
    F_P = \frac{1}{\xi D\eta_\mathrm{QE}}\left(\frac{\tau_\mathrm{tot}\left(0\right)}{\tau_\mathrm{tot}\left(F_P\right)} - 1\right)
\end{equation}
Using the experimental values of $D$ (0.25) and $\eta_\mathrm{QE}$ (0.02--0.2)  for the \thk\  site~\cite{wolfowicz2020vsq} this gives us an estimate that the Purcell factor experienced by SE1 while on resonance with the cavity is in the range of $12-118$. 
Because $\eta_\mathrm{DC}$ saturates at 1 for larger values of $F_P$ (equation \ref{eq:defect-cavity}) this large range of potential Purcell factors results in a much tighter range of defect to cavity coupling efficiencies that are largely limited by $\eta_\mathrm{CW}\times$ the waveguide to fiber coupling which is approximately $0.234$.
The Purcell factors in this range therefore yield a total efficiency of coupling for an emitted GS1-ES1 photon into fiber in the range of $0.21-0.23$. With improvements to cavity fabrication for improved cavity-to-waveguide coupling efficiency, we expect this could be raised as high as 0.33.

To try to get a tighter estimate of the Purcell enhancement we employ two additional methods to estimate the Purcell factor and compare to the range above.
Our next approach to determine the Purcell factor is via numerical simulations of the cavity's mode volume and quality factor.
Using the Tidy3D simulations discussed in Section~\ref{SIsec:Design}, we find that a defect at a depth of 150\,nm, directly under the field strength maximum of a cavity that closely matches the design of the ones used for SE1, would be expected to experience estimated Purcell factor of $F_{P}\approx 135$. 
However, the Q of this simulated cavity is $\sim3.5\times10^4$ which is substantially higher than the measured $7.9\times10^3$ observed for the cavity that we use for coupling to SE1.
Taking into account this lower Q we expect a proportionally lower Purcell enhancement of SE1, yielding $F_{P}\approx 30$.
This estimate falls well within range of Purcell factors estimated from the lifetime reduction and yeilds an estimated GS1-ES1 photon collection efficiency into fiber of 0.227.

Our final estimate of the Purcell enhancement experienced by SE1 comes from a comparison to the enhancement experienced by the defect ensemble on Sample~A.
Using comparisons of the brightness ratio between the ZPL and PSB when on and off resonance we found that the average Purcell enhancement of a V ensemble in a $Q = 12340$ cavity was upper bounded by $F_P \leq 38$ (see Section~\ref{SIsec:ensemble-lifetimes}). 
The measured optical lifetime for SE1 (109\,ns) is shorter than the stretched exponential lifetime extracted by fitting the entire ensemble in Sample A (122\,ns).
This indicates that SE1 is most likely experiencing a Purcell enhancement greater than the average of the ensemble in Sample A despite the lower Q factor of the cavity, which might be explained by strong co-location of SE1 and the cavity field maxima.
While it is difficult to extract a numerical estimate of the Purcell factor achieved for SE1 from this comparison, it does indicate a likely value on the same order of magnitude as the previous estimates. 
Due to the agreement of these three estimates we conclude that the Purcell factor is likely sufficient to nearly saturate the cavity-to-waveguide coupling efficiency, yielding a total GS1-ES1 emission to fiber coupling efficiency of at least 0.22.

When considering the overall detection efficiency in our measurements it is also important to consider losses between coupling into fiber and the registration of a photon detection event.
In Table~\ref{tab:efficiencies}, we list the transmission of each component in the optical path between the lensed fiber tip and the single photon detector as well as the estimated efficiency of the detector at our set bias current.
The total detection efficiency of a photon coupled into the fiber is estimated to be 0.0108, which when combined with the efficiency of coupling ZPL photons into the fiber discussed above yields an overall detection efficiency of ZPL photons of around 0.002.
During our $g^2(\tau)$ measurements we observed an average photon number per probe pulse of $5\times10^{-4}$ which indicates that when the cavity is on resonance, the product of the probability of defect excitation during the probe pulse and the emission probability into the GS1-ES1 ZPL ($P_\mathrm{GS1-ES1}$ in Table~\ref{tab:efficiencies}) should be approximately 0.25.

\subsection{Photon collection efficiency of confocal microscopy}
\label{SIsubsec:confocal}

In comparison to our Si-on-SiC device platform, we consider the case of defects situated in bulk SiC optically addressed using confocal microscopy.
The maximum collection efficiency of photons from these defects will be limited by the fraction of photons that can escape from the material due to its high refractive index and the numerical aperture (NA) of the microscope objective used for collection. If we first assume that we have a way to collect all of the photons that escape the material, then the theoretical limit on the collection efficiency is set by total internal reflection.
By Snell's law, the critical angle beyond which photons cannot escape a material with index $n_1$ into a second material with index $n_2 < n_1$ is given by:
\begin{equation}
    \theta_C = \mathrm{sin}^{-1}(n_2/n_1)
\end{equation}
For SiC ($n_{SiC} = 2.6$ at 1278\,nm) and vacuum ($n_{vac} = 1$) the critical angle is $\theta_C = 22.6$\textdegree.
For an isotropic point source, the fraction of emitted photons contained within a cone of angle $2\theta_C$ is the ratio of the underlying solid angle to that of the full sphere::
\begin{equation}
    \label{eq:solid-angle}
    \eta = \mathrm{sin}^2\left(\frac{\theta_C}{2}\right)
\end{equation}
This yields a photon escape fraction of 3.85\,\% from the SiC substrate, already nearly an order of magnitude lower than what was achieved in this work.

The collection efficiency is further limited when using a real objective lens.
Previous confocal microscope measurements of single \Vo:SiC defects employed microscope objectives with NA values of 0.65~\cite{wolfowicz2020vsq} and 0.85~\cite{cilibrizzi2023uni}. 
Using the formula $\mathrm{NA} = n\mathrm{sin}(\theta)$, where $\theta$ is the maximum angle the of light the objective will accept, we find that the acceptance angles through SiC for the previous experiments would be $14.5$\textdegree\ and $19.1$\textdegree\ respectively.
Taking the larger of the two NA values and replacing the critical angle in equation~\ref{eq:solid-angle} with the objective acceptance angle, we find a maximum collection efficiency of 2.75\,\%.
The collection efficiency is also reduced by the transmission of such an objective, which for the 0.85 NA objective used by Ref.~\cite{cilibrizzi2023uni} is $\sim0.68$ at 1279\,nm. This leads to a maximum collection efficiency of photons from bulk SiC of approximately 1.9\,\% meaning our platform represents an order of magnitude improvement in collection efficiency from a confocal microscope.

\section{Single Emitter Characterization}
\subsection{Theory of pulsed second-order autocorrelation on a single detector with finite dead time}
\label{SIsubsec:g2-theory}

We derive, from first principles, the expected zero-delay value of the pulsed second-order correlation function $g^{(2)}(0)$ measured on a
\emph{single} superconducting-nanowire single-photon detector (SNSPD) whose recovery (dead) time $\tau_\mathrm{d}$ is not negligible compared with the emitter lifetime $\tau_\mathrm{SE}$. 
We treat (i) a single quantum emitter, (ii) the addition of an uncorrelated background, and (iii) the two-emitter reference case that sets the multi-photon-emission threshold. 
The central result is that the usual anti-bunching benchmark $g^{(2)}(0)=1/2$ that corresponds to the resulting correlation of two equally bright single emitters with minimal background contributions is modified to $\tfrac12\mathcal{S}_\mathrm{SE}$, where $\mathcal{S}_\mathrm{SE}\le1$ is correction factor related to the dead time.

\subsubsection{Definitions and notation}
\label{SIsec:defs}

Excitation is pulsed with repetition period
\begin{equation}
    T_{\mathrm{rep}} \;=\; \tau_{\mathrm{exc}} + \tau_{\mathrm{delay}} + \tau_\mathrm{col} + \tau_{\mathrm{wait}},
    \label{eq:trep}
\end{equation}
where $\tau_{\mathrm{exc}}$ is the excitation-pulse duration, $\tau_{\mathrm{delay}}$ the delay between excitation and the opening of the collection window, $\tau_\mathrm{col}$ the collection window duration, and $\tau_{\mathrm{wait}}$ the interval between the end of the collection and the beginning of the next excitation. 
Excitations and subsequent collections are indexed by an integer probe-sequence label $k=1,\dots,P$, and the acquisition comprises $N_{\mathrm{cyc}}$ independent repump-probe sequence cycles, in the beginning of which the charge state of the single emitter candidate is reset by a single charge repump sequence. 
Coincidences are binned by the integer probe-sequence index difference $m$ between probe-sequences $k$ and $k+m$, each of which can contain an arbitrary number of detection events.

\paragraph{Detected-count means.}
Let $\bar\mu$ be the mean number of \emph{detected} emitter photons per excitation and $\bar\beta$ the mean number of uncorrelated background counts (predominantly detector dark counts, as well as some stray light) per collection window. 
The isolated detection efficiency $\eta_0$ of the SNSPD is common to all collected photons and is absorbed into $\bar\mu$ from the outset; similarly for $\bar\beta$. 
Both means are dimensionless and small, $\bar\mu,\bar\beta\ll1$, in the regime of interest. 
We define the signal-to-background ratio $\rho$ and the signal fraction $\sigma$ as
\begin{equation}
    \rho \equiv \frac{\bar\mu}{\bar\beta}, \qquad
    \sigma \equiv \frac{\bar\mu}{\bar\mu+\bar\beta}
     = \frac{\rho}{1+\rho}\in[0,1].
    \label{eq:sigma}
\end{equation}

\paragraph{The measured statistic.}
The detector is non-photon-number-resolving: within one collection
window it registers a number of counts $n_k\in\{0,1,2,\dots\}$, each an event of assumed single-photon or single dark-count origin. 
The zero-delay coincidence count is the number of \emph{ordered} (pair of events is counted twice) within-window count pairs,
\begin{equation}
    C_0 \;=\; \sum_{k}\, n_k\,(n_k-1).
    \label{eq:c0def}
\end{equation}
More generally, the coincidence count at integer delay $m\neq0$ counts ordered pairs of counts in probe sequences separated by $m$,
\begin{equation}
    C_m \;=\; \sum_{k}\, n_k\,n_{k+m}.
    \label{eq:cmdef}
\end{equation}
For $|m|$ large enough that the two probe sequences are statistically independent, $C_m$ tends to a constant uncorrelated level, denoted $C_\infty$, which normalizes the correlation,
\begin{equation}
    g^{(2)}(m) \;=\; \frac{C_m}{C_\infty}.
    \label{eq:g2def}
\end{equation}
The construction of $C_\infty$ is given in Sec.~\ref{SIsec:-tail}.
Uncertainties on $g^{(2)}(m)$ are estimated model-free from the cycle-to-cycle spread (Sec.~\ref{SIsec:errors}), not assuming Poisson statistics for the coincidence counts.

\subsubsection{Detection event and detector models}
\label{SIsec:-models}

\paragraph{Emitter.}
Under excitation from a single pulse, each emitter contributes at most one photon per excitation (a Bernoulli event), with a mean number of $\bar\mu$  per window.
The spontaneous-emission time, measured from excitation, is exponentially distributed with lifetime $\tau_\mathrm{SE}$; we use $\lambda\equiv1/\tau_\mathrm{SE}$ for brevity. 
The collection window (gate) opens at $t_0=\tau_{\mathrm{exc}}+\tau_{\mathrm{delay}}$ and closes at $t_0+\tau_\mathrm{col}$. 
Using $\theta=t-t_0$ for the time since the gate opens, the \emph{detected} (gated and renormalized) emission arrival-time density is the truncated exponential
\begin{equation}
    p_\mathrm{SE}(\theta) \;=\;
    \frac{\lambda\,e^{-\lambda\theta}}{1-e^{-\lambda\tau_\mathrm{col}}},
    \qquad 0\le\theta\le\tau_\mathrm{col}, \qquad
    Z\equiv 1-e^{-\lambda\tau_\mathrm{col}}.
    \label{eq:pse-trunc}
\end{equation}
The gate delay $t_0$ shifts all photons of a given period equally and therefore cancels in any time-difference; only the finite window $\tau_\mathrm{col}$ has an observable effect on correlations, through the normalization factor $Z$.

\paragraph{Background.}
Background counts are treated as a homogeneous Poisson process of independent and uncorrelated detection events over the gate, with mean number $\bar\beta$ per window and uniform arrival density
\begin{equation}
    p_\mathrm{B}(\theta) \;=\; \frac{1}{\tau_\mathrm{col}},
    \qquad 0\le\theta\le\tau_\mathrm{col},
\label{eq:pB}
\end{equation}

\paragraph{Detector recovery.}
After a count at time $t_c$, the detection efficiency for a subsequent photon at $t>t_c$ is reduced by a generic normalized recovery function $R(\Delta)$ of the separation $\Delta=t-t_c$,
\begin{equation}
    \eta_{\mathrm{eff}}(t\,|\,\text{count at }t_c)
     \;=\; \eta_0\,R(t-t_c),
    \qquad R(0^+)\approx0,\quad R(\infty)=1.
    \label{eq:Rdef}
\end{equation}
A detection resets the recovery clock.
We consider two recovery function forms,
\begin{align}
    R_{\mathrm{scr}}(\Delta) &= \frac{1+\erf\left[C\left(1-e^{-\left(\Delta - \tau_\mathrm{on}\right)/\tau_\mathrm{scc}}\right)\right]}{1+\erf\left[C\right]}
     &&\text{(SNSPD supercurrent recovery),}
    \label{eq:Rscr}\\
    R_{\mathrm{step}}(\Delta) &= \Theta(\Delta-\tau_\mathrm{step})
    &&\text{(unit step approximation).}
    \label{eq:Rstep}
\end{align}
The error function form models the normalized Quantum Opus One SNSPD efficiency as a function of bias-current recovery on the circuit timescale $\tau_\mathrm{scc}$, with $C$ and $\tau_\mathrm{on}$ being related to the bias current.
Here $\tau_\mathrm{on}$ marks the recovery onset (where the erf argument vanishes) and $\tau_\mathrm{scc}$ the soft-edge width; the recovery has an effective rise time of order $\tau_\mathrm{scc}/C$.
For either $\tau_\mathrm{scc}$ small or $C$ large, $R_\mathrm{scr}$ transitions from $0$ to $1$ over a range much shorter than $\tau_\mathrm{on}$, approaching a unit step at $\Delta=\tau_\mathrm{on}$.
In this regime the step form $R_\mathrm{step}$ captures the recovery with a single parameter $\tau_\mathrm{step}$ and yields closed-form coincidence integrals.
We therefore use $R_\mathrm{step}$ for analytic results, which we validate by comparing against leading and subleading order approximations of any $R$ generally or $R_\mathrm{scr}$ specifically. 
When the sharp-transition approximation is not appropriate, one could evaluate any physical recovery model $R$, including $R_\mathrm{scr}$, numerically.

Whatever the recovery shape, the coincidence rates below depend on $R$ primarily through low moments of the \emph{recovery pdf} $R'(s)$: since $R$ grows monotonically from $0$ to $1$, $R'$ is a normalized density (up to $\mathcal{O}(1-\mathrm{erf}\,C)$ for $R_\mathrm{scr}$ from the finite erf plateau).
Integration by parts gives the identity
\begin{equation}
M_n \equiv \int_0^\infty s^n\,[1-R(s)]\,ds \;=\; \frac{\langle s^{n+1}\rangle_{R'}}{n+1},
\label{eq:Mn-identity}
\end{equation}
so all deficit moments are moments of $R'$.
In particular the \emph{effective dead time}
\begin{equation}
\tau_\mathrm{d}\;\equiv\;\int_0^{\infty}\!\left[1-R(s)\right]\,ds \;=\; \langle s\rangle_{R'}
\label{eq:taud}
\end{equation}
is the mean recovery time (the zeroth moment of the deficit $1-R$, or the first moment of $R'$), and the next-order shape of $R$ is captured by the variance
\begin{equation}
\sigma_R^2 \equiv \langle s^2\rangle_{R'} - \tau_\mathrm{d}^2 = 2M_1 - \tau_\mathrm{d}^2.
\label{eq:sigmaR}
\end{equation}
Equivalently, $\tau_\mathrm{d}$ is the location of $R$ conserving the total deficit: the area kept in error below it, $\int_0^{\tau_\mathrm{d}}R\,ds$, equals the area dropped in error above it, $\int_{\tau_\mathrm{d}}^{\infty}[1-R]\,ds$.
For the step model $R'=\delta(s-\tau_\mathrm{step})$, giving $\tau_\mathrm{d}^\mathrm{step}=\tau_\mathrm{step}$ and $\sigma_R^\mathrm{step}=0$.
For $R_\mathrm{scr}$, using  $x=C(1-e^{-(s-\tau_\mathrm{on})/\tau_\mathrm{scc}})$ maps $R_\mathrm{scr}'(s)\,ds$ onto a truncated Gaussian $\mathcal{G}(x)=\tfrac{2}{\sqrt{\pi}[1+\mathrm{erf}\,C]}e^{-x^2}$ on $x\in(-\infty,C]$, peaked at $s=\tau_\mathrm{on}$ with width $\sim\tau_\mathrm{scc}/C$; expanding $s(x)=\tau_\mathrm{on}+\tau_\mathrm{scc}\sum_{n\ge 1}x^n/(nC^n)$ and taking Gaussian moments (extending the upper limit $C\to\infty$, error $\mathcal{O}(1-\mathrm{erf}\,C)$) yields
\begin{equation}
\tau_\mathrm{d}^\mathrm{scr} = \tau_\mathrm{on} + \frac{\tau_\mathrm{scc}}{2C^2[1+\mathrm{erf}\,C]}\left[1+\frac{3}{4C^2}+\ldots\right],
\qquad
\sigma_R^{2,\mathrm{scr}} = \frac{\tau_\mathrm{scc}^2}{C^2[1+\mathrm{erf}\,C]}+\mathcal{O}\!\left(\frac{\tau_\mathrm{scc}^2}{C^4}\right),
\label{eq:taud-sigma-scr}
\end{equation}
both reducing to the step values ($\tau_\mathrm{on},\,0$) in the sharp-transition limit ($C\to\infty$ or $\tau_\mathrm{scc}/\tau_\mathrm{on}\to 0$). 
The positive shift $\tau_\mathrm{d}^\mathrm{scr}-\tau_\mathrm{on}$ reflects the asymmetry of $s(x)$, which stretches more above $\tau_\mathrm{on}$ than below.


\subsubsection{Uncorrelated level and normalization}
\label{SIsec:-tail}

The denominator $C_\infty$ in Eq.~\eqref{eq:g2def} is the coincidence level between two probe sequences so far apart that they are statistically independent, and normalizing by it converts a raw coincidence count into $g^{(2)}$. 
Within a single repump--probe cycle, however, consecutive probe sequences are \emph{not} independent:
(i) Ionization removes the emitter irreversibly for the remainder of a cycle, which would appear as an average decay of the mean rate across the probe index; 
we suppress it by choosing a resonant excitation power at which ionization is slow and negligible over the probe sequences used in each cycle. 
(ii) The repump at the beginning of each cycle resets the charge state and causes a deterministic drift of the mean detected rate across the probe index within a cycle;
we minimize its effect by discarding the initial probe sequences over which this transient is appreciable. 
(iii) Spectral diffusion causes the emitter to drift into and out of resonance with the excitation laser; 
while near resonance the detected rate briefly rises, producing a positive (bunching) correlation between sequences $k$ and $k+m$ that is largest at small $|m|$ and decays to a constant as $|m|\to\infty$. 
After (i) and (ii) the retained mean profile $\langle n_k\rangle$ is flat in $k$, and the only residual correlation, spectral diffusion, vanishes at large separation. 
We therefore define the uncorrelated normalization $C_\infty$ as the coincidence level in the large-$|m|$ limit, where the sequences decorrelate.
Taking the expectation of Eq.~\eqref{eq:cmdef} for $m\neq0$,
\begin{equation}
\big\langle C_m\big\rangle
= \sum_k \big\langle n_k\,n_{k+m}\big\rangle
\;\xrightarrow{}\;
\lim_{\left|m\right|\to\infty}\big\langle C_m\big\rangle=\sum_k \big\langle n_k\big\rangle\big\langle n_{k+m}\big\rangle .
\label{eq:cm-expect}
\end{equation}

With the retained mean profile flat, $\langle n_k\rangle=\bar\nu$ for all $k$, where $\bar\nu\equiv\bar\mu+\bar\beta$ is the ensemble-mean detected counts per probe sequence. 
To populate every coincidence bin $m\in[-W,+W]$ (with $W$ the number of probe-sequences that represent the half-width of the coincidence histogram) from a common set of reference sequences, only interior sequences possessing all neighbours out to $\pm W$ are used as
references. 
There are $P_\mathrm{eff}=P-2W$ such sequences, and each contributes once to every bin, so the histogram is flat across $|m|\le W$. 
Summing Eq.~\eqref{eq:cm-expect} over these $P_\mathrm{eff}$ reference sequences and over the $N_{\mathrm{cyc}}$ cycles,
\begin{equation}
\big\langle C_\infty\big\rangle
= N_{\mathrm{cyc}}\,P_\mathrm{eff}\,\bar\nu^{\,2}.
\label{eq:Cinf-mean}
\end{equation}

Because the emitter brightness may vary from cycle to cycle, we build $C_\infty$ as a sum of per-cycle contributions, each using that cycle's own mean rate. 
A natural per-cycle rate estimator is $\hat{\bar\nu}_c \equiv K_c^{\mathrm{int}}/P_\mathrm{eff}$, but its square $\hat{\bar\nu}_c^{\,2}$ is biased high as an estimator of $\bar\nu_c^{\,2}$ because expanding $(K_c^{\mathrm{int}})^2$ includes same-pulse (diagonal) terms $n_{c,k}^2$, whose expectation contains the shot-noise variance in addition to $\bar\nu_c^{\,2}$. 
We therefore cannot use it directly and instead construct an unbiased estimator by removing the diagonal.
The uncorrelated coincidence level at nonzero delay pairs an \emph{interior reference} pulse with \emph{any partner} pulse within the histogram half-width. 
The relevant statistic is thus the sum over ordered interior-reference/partner pairs with distinct indices,
\begin{equation}
S_c=\sum_{k\in\mathrm{int}}\sum_{j\ne k}n_{c,k}\,n_{c,j}
=K_c^{\mathrm{int}}K_c-\sum_{k\in\mathrm{int}}n_{c,k}^2 ,
\end{equation}
where $K_c=\sum_k n_{c,k}$ and $K_c^{\mathrm{int}}=\sum_{k\in\mathrm{int}}n_{c,k}$. Only interior self-pairs are subtracted, since edge pulses never serve as references. Under a flat, uncorrelated profile, each of the $P_\mathrm{eff}(P-1)$ distinct ordered pairs has expectation $\bar\nu_c^{\,2}$, giving the unbiased estimator
\begin{equation}
\widehat{\bar\nu_c^{\,2}}=\frac{K_c^{\mathrm{int}}K_c-\sum_{k\in\mathrm{int}}n_{c,k}^2}{P_\mathrm{eff}(P-1)},
\qquad \big\langle\widehat{\bar\nu_c^{\,2}}\big\rangle=\bar\nu_c^{\,2}.
\label{eq:unbiased}
\end{equation}
In the flat limit this coincides with $\hat{\bar\nu}_c^{\,2}$ to $\mathcal{O}(1/P)$. The uncorrelated level is then assembled per cycle as
\begin{equation}
\boxed{\;
C_\infty=\sum_{c=1}^{N_{\mathrm{cyc}}} t_c,
\qquad
t_c\equiv P_\mathrm{eff}\,\widehat{\bar\nu_c^{\,2}}
=\frac{K_c^{\mathrm{int}}K_c-\sum_{k\in\mathrm{int}}n_{c,k}^2}{P-1}.\;}
\label{eq:tail}
\end{equation}

\paragraph{Non-uniformity of the emission profile.}
Equation~\eqref{eq:tail} assumes a flat retained mean profile. 
To certify this, we allow $\langle n_k\rangle=\bar\nu_c f_k$ with $\sum_k f_k=P$ and define the lag-resolved participation factor
\begin{equation}
\kappa_m\equiv\frac{P_\mathrm{eff}^{-1}\sum_{k\in\mathrm{int}}f_k f_{k+m}}
{\left[P_\mathrm{eff}^{-1}\sum_{k\in\mathrm{int}}f_k\right]^2},
\label{eq:kappa}
\end{equation}
with $\kappa_m\to1$ for a flat or decorrelated profile and $\kappa_0=\langle f_k^2\rangle/\langle f_k\rangle^2\ge1$. 
An uncorrected non-flat profile would inflate $g^{(2)}(0)$ by $\kappa_0$ and mimic bunching at small $m$. 
Here, we use $\kappa_m$ only as a diagnostic rather than a correction.

The estimator $\hat\kappa_m$ built from the cycle-averaged profile $\hat f_k$ is unbiased for $m\neq0$ but biased high at $m=0$ by sampling noise, sitting on a shot-noise floor
\begin{equation}
\big\langle\hat\kappa_0\big\rangle_\mathrm{flat}=1+\frac{1}{N_\mathrm{cyc}\,\bar\nu}.
\label{eq:kappa-floor}
\end{equation}
In our data, we confirm flatness with the debiased $\hat\kappa_0-1/(N_\mathrm{cyc}\bar\nu)\approx1$ and $\hat\kappa_{m\neq0}\approx1$. 
\emph{No} $\kappa$ correction is applied to the $g^{(2)}(0)$ estimate.

\subsubsection{Zero-delay coincidence for a single emitter}
\label{SIsec:-central-single}

For an ideal single emitter that cannot be excited twice in a single probe sequence, two counts in one window ($n_k\ge2$) require at least one non-emitter count.
Hence, in the absence of background, same-window coincidences are $C_0=0$; in the presence of background, all same-window coincidences of an ideal single emitter are background-driven.

The expectation of Eq.~\eqref{eq:c0def} is given by
\begin{equation}
    \langle C_0\rangle
    = \sum_{k}\big\langle n_k(n_k-1)\big\rangle
    = N_{\mathrm{cyc}}\,P_\mathrm{eff}\,\big\langle n(n-1)\big\rangle,
    \label{eq:c0-expect}
\end{equation}
where the last equality uses that each eligible probe sequence has the same statistics, so the index-independent per-pulse mean $\langle n(n-1)\rangle$ multiplies the number of eligible pulse sequences.

For rare independent events, $\langle n(n-1)\rangle$ is the expected number of ordered detected pairs in one collection window.
Writing each arrival-time density as (mean number)$\times$(normalized shape), e.g.\ $\bar\mu\,p_\mathrm{SE}$ or $\bar\beta\,p_\mathrm{B}$, the pair rate factorizes into an amplitude and a shape integral,
\begin{equation}
\big\langle n(n-1)\big\rangle
= (\text{mean}_1)(\text{mean}_2)\,\mathcal{S}[p_1,p_2],
\qquad
\mathcal{S}[p_1,p_2]\equiv\iint p_1(t_1)\,p_2(t_2)\,P_\mathrm{both}(t_1,t_2)\,dt_1\,dt_2 ,
\label{eq:pairmaster}
\end{equation}
where
\begin{equation}
    P_{\mathrm{both}}(t_1,t_2)
    \;=\; R\!\left(|t_1-t_2|\right)
    \label{eq:atom}
\end{equation}
is the probability that two candidate photons at $t_1,t_2$ are both detected on the single detector, and $\mathcal{S}$ is the pair-survival factor (the fraction of candidate pairs that survive detector recovery).
Since the detector efficiency is absorbed in both $\bar\mu$ and $\bar\beta$, $P_\mathrm{both}$ is directly the normalized recovery function $R$; because it depends only on the separation $s=|t_1-t_2|$, $\mathcal{S}$ is equivalently an average of $R(s)$ over the distribution of separations induced by $p_1,p_2$.

Working in the gate-relative time $\theta=t-t_0$ of Eq.~\eqref{eq:pse-trunc} (the offset $t_0$ cancels in $|t_1-t_2|$), and writing $\theta_1,\theta_2$ for the two count times, it is convenient to introduce the accumulated deficit
\begin{equation}
G(L)\equiv\int_0^{L}\!\left[1-R(s)\right]\,ds ,
\label{eq:Gdef}
\end{equation}
which grows from $G(0)=0$ and saturates at $G(\infty)=\tau_\mathrm{d}$ [Eq.~\eqref{eq:taud}], since $R(s)\to1$ at large $s$.
For the step model $G(L)=\min(L,\tau_\mathrm{step})$: it grows linearly until $L=\tau_\mathrm{step}$ and is flat thereafter, so $\tau_\mathrm{d}^\mathrm{step}=\tau_\mathrm{step}$.

When one count is uniform ($p_2=p_\mathrm{B}=1/\tau_\mathrm{col}$), the $\theta_2$ integral in Eq.~\eqref{eq:pairmaster} can be computed first.
For a first count at $\theta_1$, the coincident click can arrive earlier (within $[0,\theta_1]$, separations up to $\theta_1$) or later (within $[\theta_1,\tau_\mathrm{col}]$, separations up to $\tau_\mathrm{col}-\theta_1$), so the total lost coincidence is $G(\theta_1)+G(\tau_\mathrm{col}-\theta_1)$ and
\begin{equation}
\mathcal{S}_\mathrm{B}\equiv\mathcal{S}[p_1,p_\mathrm{B}]
= 1-\frac{1}{\tau_\mathrm{col}}\int_0^{\tau_\mathrm{col}}
\!\left[G(\theta_1)+G(\tau_\mathrm{col}-\theta_1)\right]\,p_1(\theta_1)\,d\theta_1 .
\label{eq:Sgen}
\end{equation}

This deficit has a model-independent leading form.
Writing $G(\theta_1)=\tau_\mathrm{d}-\varepsilon(\theta_1)$ with the tail $\varepsilon(\theta_1)=\int_{\theta_1}^{\infty}[1-R]\,ds$, and using $\int p_1\,d\theta_1=1$,
\begin{equation}
\mathcal{S}_\mathrm{B}
= 1-\frac{2\tau_\mathrm{d}}{\tau_\mathrm{col}}
+\frac{1}{\tau_\mathrm{col}}\int_0^{\tau_\mathrm{col}}\!\left[\varepsilon(\theta_1)+\varepsilon(\tau_\mathrm{col}-\theta_1)\right]\,p_1(\theta_1)\,d\theta_1 ,
\label{eq:Sexact}
\end{equation}
giving
\begin{equation}
\boxed{\;
\mathcal{S}_\mathrm{B}=1-\frac{2\tau_\mathrm{d}}{\tau_\mathrm{col}}+\mathcal{O}\!\left(\frac{\tau_\mathrm{d}^2}{\tau_\mathrm{col}^2}\right),\;}
\label{eq:Sleading}
\end{equation}
independent of the first-count density $p_1$ and of the recovery model at leading order.
The subleading correction is set by the tail integral of $\varepsilon$, which by swapping the order of integration is $\int_0^\infty\varepsilon(\theta)\,d\theta=M_1=\tfrac12(\tau_\mathrm{d}^2+\sigma_R^2)$; the specific $p_1$ then determines how this moment enters.

To leading order in $\bar\mu,\bar\beta$, two channels contribute to the zero-delay coincidence: (i) two background counts, and (ii) one single-emitter photon and one background count.

\paragraph{Channel 2B (two background counts).}
Both counts are uniform, $p_1=p_2=p_\mathrm{B}$, so Eq.~\eqref{eq:pairmaster} gives
\begin{equation}
\big\langle n(n-1)\big\rangle_{\mathrm{2B}}
= \bar\beta^{\,2}\,\mathcal{S}_\mathrm{B}^{\mathrm{2B}},
\qquad
\mathcal{S}_\mathrm{B}^{\mathrm{2B}}\equiv\mathcal{S}[p_\mathrm{B},p_\mathrm{B}].
\label{eq:chanB}
\end{equation}
With $p_1=1/\tau_\mathrm{col}$ uniform, the $\varepsilon$ integral in Eq.~\eqref{eq:Sexact} yields $M_1/\tau_\mathrm{col}$ from each endpoint, so
\begin{equation}
\mathcal{S}_\mathrm{B}^{\mathrm{2B}} \approx 1 - \frac{2\tau_\mathrm{d}}{\tau_\mathrm{col}} + \frac{\tau_\mathrm{d}^2+\sigma_R^2}{\tau_\mathrm{col}^2}.
\label{eq:SB-2B-approx}
\end{equation}
For the step model ($\sigma_R=0$) this is also the exact solution,
\begin{equation}
\mathcal{S}_\mathrm{B}^{\mathrm{2B,step}}=\left(1-\frac{\tau_\mathrm{step}}{\tau_\mathrm{col}}\right)^2
=1-\frac{2\tau_\mathrm{step}}{\tau_\mathrm{col}}+\frac{\tau_\mathrm{step}^2}{\tau_\mathrm{col}^2}.
\label{eq:SB-2B}
\end{equation}

\paragraph{Channel 1SE\,$+$\,B (emitter photon \& one background count).}
Now the first count carries the emitter density $p_1=p_\mathrm{SE}$ [Eq.~\eqref{eq:pse-trunc}] and the second is the uniform background $p_2=p_\mathrm{B}$.
Ordered-pair counting (a detected pair increments $n(n-1)$ by $2$) gives
\begin{equation}
\big\langle n(n-1)\big\rangle_{\mathrm{1SE+B}}
= 2\,\bar\mu\,\bar\beta\,\mathcal{S}_\mathrm{B}^{\mathrm{1SE+B}},
\qquad
\mathcal{S}_\mathrm{B}^{\mathrm{1SE+B}}\equiv\mathcal{S}[p_\mathrm{SE},p_\mathrm{B}].
\label{eq:chanA}
\end{equation}
Because $p_\mathrm{SE}$ is concentrated near $\theta=0$, only the $\varepsilon(\theta_1)$ contribution matters (the mirror $\varepsilon(\tau_\mathrm{col}-\theta_1)\to\varepsilon(\tau_\mathrm{col})\to 0$); integrating by parts with the emitter CDF $P_\mathrm{SE}(\theta)=(1-e^{-\lambda\theta})/Z$ converts $\int_0^\infty\varepsilon\,p_\mathrm{SE}\,d\theta$ into $[\tau_\mathrm{d}-\widetilde{D}(\lambda)]/Z$ with $\widetilde{D}(\lambda)=\int_0^\infty e^{-\lambda\theta}[1-R]\,d\theta = \tau_\mathrm{d}-\lambda M_1+\ldots$, so
\begin{equation}
\mathcal{S}_\mathrm{B}^{\mathrm{1SE+B}}
\approx 1 - \frac{2\tau_\mathrm{d}}{\tau_\mathrm{col}} + \frac{\lambda M_1}{Z\,\tau_\mathrm{col}} + \ldots
\;\xrightarrow{\tau_\mathrm{col}\gg\tau_\mathrm{SE}}\;
1 - \frac{2\tau_\mathrm{d}}{\tau_\mathrm{col}} + \frac{\tau_\mathrm{d}^2+\sigma_R^2}{2\tau_\mathrm{SE}\,\tau_\mathrm{col}}.
\label{eq:SB-1SEB-approx}
\end{equation}
Evaluating Eq.~\eqref{eq:Sgen} with $p_1=p_\mathrm{SE}$ for the step model yields the closed form
\begin{equation}
\mathcal{S}_\mathrm{B}^{\mathrm{1SE+B,step}}=1-
\frac{\lambda\tau_\mathrm{step}
+\sinh\left(\lambda\tau_\mathrm{step}\right)
+\left(1-\cosh\left(\lambda\tau_\mathrm{step}\right)\right)\coth\left(\tfrac{\lambda\tau_\mathrm{col}}{2}\right)
}{\lambda\tau_\mathrm{col}},
\label{eq:SB-1SEB}
\end{equation}
which for $\tau_\mathrm{col}\gg\tau_\mathrm{step}$ reduces to $1-2\tau_\mathrm{step}/\tau_\mathrm{col}+\tau_\mathrm{step}^2/(2\tau_\mathrm{SE}\tau_\mathrm{col})+\ldots$, matching the general result with $\sigma_R=0$.
We therefore see that generally $\mathcal{S}_\mathrm{B}^\mathrm{scr}>\mathcal{S}_\mathrm{B}^\mathrm{step}$ at fixed $\tau_\mathrm{d}$, meaning that using the step function approximation is a good conservative limit.

\paragraph{Result.}
Combining Eqs.~\eqref{eq:c0-expect}, \eqref{eq:chanB} and~\eqref{eq:chanA}, the single-emitter-plus-background zero-delay coincidence is
\begin{equation}
\boxed{\;
C_0^{(\mathrm{1SE+B;2B})}
\;=\; N_{\mathrm{cyc}}\,P_\mathrm{eff}\,
\left(2\bar\mu\bar\beta\,\mathcal{S}_\mathrm{B}^{\mathrm{1SE+B}}
+\bar\beta^{\,2}\,\mathcal{S}_\mathrm{B}^{\mathrm{2B}}\right)
\;+\;\mathcal{O}(\bar\mu\bar\beta^{2},\bar\beta^{3}).\;}
\label{eq:c0-single}
\end{equation}
Both channels share the common leading deficit $1-2\tau_\mathrm{d}/\tau_\mathrm{col}$ and differ only in the subleading edge correction.
Because emitter and background (or two background) counts are distributed over the full window $\tau_\mathrm{col}\gg\tau_\mathrm{d}$, the dead-time suppression is weak, $\mathcal{S}_\mathrm{B}^\mathrm{1SE+B}\approx\mathcal{S}^\mathrm{2B}_\mathrm{B}\approx\mathcal{S}_\mathrm{B}\approx1-2\tau_\mathrm{d}/\tau_\mathrm{col}$.
This is qualitatively different from the two-emitter case (Sec.~\ref{SIsec:-two}), where the paired photons are bunched within $\sim\tau_\mathrm{SE}\sim\tau_\mathrm{d}$ and the suppression $\mathcal{S}_\mathrm{SE}$ is stronger.

\subsubsection{Two emitters: the dead-time-shifted threshold}
\label{SIsec:-two}

To calibrate the multi-photon-emission threshold we consider two independent emitters of the same lifetime $\tau_\mathrm{SE}$, with detected means $\bar\mu_1$ and $\bar\mu_2=r\,\bar\mu_1$, relative brightness $r\in[0,1]$, and total $\bar\mu=\bar\mu_1+\bar\mu_2$.
A same-window coincidence now requires both emitters to fire in the same period.

\paragraph{Emitter pair-survival factor.}
Both emission times $\theta_1,\theta_2$ are independent draws from $p_\mathrm{SE}$ [Eq.~\eqref{eq:pse-trunc}], so this channel is the pair-survival factor of Eq.~\eqref{eq:pairmaster} with both densities equal to $p_\mathrm{SE}$.
Ordered-pair counting gives
\begin{equation}
\big\langle n(n-1)\big\rangle_{\mathrm{2SE}}
= 2\,\bar\mu_1\bar\mu_2\,\mathcal{S}_\mathrm{SE},
\qquad
\mathcal{S}_\mathrm{SE}\equiv\mathcal{S}[p_\mathrm{SE},p_\mathrm{SE}].
\label{eq:2SE}
\end{equation}
Unlike the background channels, where one count is uniform and the deficit reduces to $G(\theta_1)+G(\tau_\mathrm{col}-\theta_1)$ [Eq.~\eqref{eq:Sgen}], here both counts follow $p_\mathrm{SE}$; it is then simplest to average $R(s)$ directly over the distribution of the emission separation $s=|\theta_1-\theta_2|$,
\begin{equation}
\mathcal{S}_\mathrm{SE}
= \int_0^{\tau_\mathrm{col}} p_\Delta(s)\,R(s)\,ds ,
\label{eq:RSE-general}
\end{equation}
with $p_\Delta(s)$ the emission-separation density.
Since $\theta_1,\theta_2$ are i.i.d.\ from Eq.~\eqref{eq:pse-trunc}, the density of $s$ is obtained by fixing the smaller time as $\theta$, integrating over its allowed range, and multiplying by $2$ for the two orderings,
\begin{align}
p_\Delta(s)
&= 2\int_0^{\tau_\mathrm{col}-s} p_\mathrm{SE}(\theta)\,p_\mathrm{SE}(\theta+s)\,d\theta
= \frac{2\lambda^2 e^{-\lambda s}}{Z^2}
 \int_0^{\tau_\mathrm{col}-s} e^{-2\lambda\theta}\,d\theta
\nonumber\\[0.5ex]
&= \frac{\lambda\,e^{-\lambda s}}{Z^2}
 \left(1-e^{-2\lambda(\tau_\mathrm{col}-s)}\right),
\qquad 0\le s\le\tau_\mathrm{col}.
\label{eq:sep-trunc}
\end{align}
One verifies $\int_0^{\tau_\mathrm{col}}p_\Delta(s)\,ds=1$, and as $\tau_\mathrm{col}\to\infty$ ($Z\to1$) it reduces to the folded-exponential form $p_\Delta(s)\to\lambda e^{-\lambda s}$.

Writing $R=1-(1-R)$ and splitting $p_\Delta$ into its two exponential pieces, the deficit contribution to $\mathcal{S}_\mathrm{SE}$ becomes a sum of moments $M_n=\int_0^\infty s^n[1-R]\,ds$ (extending the upper limit costs only $\mathcal{O}(e^{-\tau_\mathrm{col}/\tau_\mathrm{d}})$, since $1-R$ dies well before the window edge).
Truncating at $M_1$ and reusing $M_0=\tau_\mathrm{d}$, $M_1=\tfrac12(\tau_\mathrm{d}^2+\sigma_R^2)$ from Eqs.~\eqref{eq:taud},~\eqref{eq:sigmaR},
\begin{equation}
\mathcal{S}_\mathrm{SE}
= 1 - \coth\!\left(\frac{\lambda\tau_\mathrm{col}}{2}\right)\lambda\tau_\mathrm{d}
+ \frac{\cosh(\lambda\tau_\mathrm{col})}{\cosh(\lambda\tau_\mathrm{col})-1}\cdot\frac{\lambda^2(\tau_\mathrm{d}^2+\sigma_R^2)}{2}
+ \mathcal{O}(\lambda^3 M_2),
\label{eq:SSE-general}
\end{equation}
where the hyperbolic factors are the finite-window edge enhancements at $M_0$ and $M_1$ order (both $\to 1$ as $\tau_\mathrm{col}\to\infty$).
For a long window this reduces into the compact form
\begin{equation}
\boxed{\;
\mathcal{S}_\mathrm{SE}^\infty \approx \exp\!\left[-\frac{\tau_\mathrm{d}}{\tau_\mathrm{SE}} + \frac{\sigma_R^2}{2\tau_\mathrm{SE}^2}\right],\;}
\label{eq:SSE-inf}
\end{equation}
exact at $\mathcal{O}(\lambda^2)$ and correct at $\mathcal{O}(\lambda^3)$ up to the third central moment (skewness) of $R'$; finite-window corrections are of order $e^{-\tau_\mathrm{col}/\tau_\mathrm{SE}}$.

Inserting $p_\Delta$ into Eq.~\eqref{eq:RSE-general} with $R_\mathrm{step}$ [Eq.~\eqref{eq:Rstep}] gives the exact closed form for $\tau_\mathrm{d}<\tau_\mathrm{col}$,
\begin{equation}
\mathcal{S}_\mathrm{SE}^{\mathrm{step}}
= \frac{\cosh{\left(\lambda\left(\tau_\mathrm{col}-\tau_\mathrm{d}\right)\right)}-1}{\cosh{\left(\lambda\tau_\mathrm{col}\right)}-1},
\label{eq:RSE-step}
\end{equation}
which reproduces Eq.~\eqref{eq:SSE-general} to $\mathcal{O}(\lambda^2\tau_\mathrm{d}^2)$ when $\sigma_R\to 0$, and reduces to
\begin{equation}
\mathcal{S}_\mathrm{SE}^{\mathrm{step},\infty}=e^{-\tau_\mathrm{d}/\tau_\mathrm{SE}}
\label{eq:RSE-inf}
\end{equation}
for $\tau_\mathrm{col}\gg\tau_\mathrm{SE}$, which is also the $\sigma_R=0$ limit of Eq.~\eqref{eq:SSE-inf}.
From Eqs. (\ref{eq:SSE-general} -- \ref{eq:RSE-step}) we see that generally $\mathcal{S}_\mathrm{SE}^\mathrm{scr}>\mathcal{S}_\mathrm{SE}^\mathrm{step}$ at fixed $\tau_\mathrm{d}$.
Therefore, similar to $\mathcal{S}_\mathrm{B}$, using the step function approximation for $\mathcal{S}_\mathrm{SE}$ is a good conservative limit.

\paragraph{Two-emitter zero-delay, $g^{(2)}(0)$, and threshold.}
Following Eq.~\eqref{eq:c0-expect}, the two-emitter zero-delay coincidence is
\begin{equation}
C_0^{(\mathrm{2SE})}
= N_{\mathrm{cyc}}\,P_\mathrm{eff}\times
 2\,\bar\mu_1\bar\mu_2\,\mathcal{S}_\mathrm{SE}.
\label{eq:c0-2SE}
\end{equation}
Combining with the level $C_\infty=N_{\mathrm{cyc}}P_\mathrm{eff}(\bar\mu_1+\bar\mu_2)^2$ and using $\bar\mu_1\bar\mu_2=r\,\bar\mu^{\,2}/(1+r)^2$,
\begin{equation}
\boxed{\;
g^{(2)}_{\mathrm{2SE}}(0)=\frac{2r}{(1+r)^2}\,\mathcal{S}_\mathrm{SE}.\;}
\label{eq:g2-2SE}
\end{equation}
The maximum over $r\in[0,1]$ occurs at $r=1$ (two equally bright emitters) and defines the threshold that replaces the classical value $1/2$,
\begin{equation}
\boxed{\;
g^{(2)}_{\mathrm{thr}}
=g^{(2)}_{\mathrm{2SE}}(0)\big|_{r=1}
=\tfrac12\,\mathcal{S}_\mathrm{SE}\;<\;\tfrac12 .\;}
\label{eq:threshold}
\end{equation}
Dead time preferentially suppresses the closely spaced same-window pairs that constitute the true two-emitter signal, while leaving the uncorrelated level unaffected; the two-emitter benchmark is thus pulled \emph{below} $1/2$.
More generally, for $N$ equally bright emitters, $g^{(2)}(0)=(1-1/N)\,\mathcal{S}_\mathrm{SE}$.

\subsubsection{Background admixture: measured versus corrected \texorpdfstring{$g^{(2)}(0)$}{g2(0)}}
\label{SIsec:-bg}

With both a second emitter (brightness $r$), background present and assuming $\tau_\mathrm{col}\gg\tau_\mathrm{d}$ and $\tau_\mathrm{col}\gg\tau_\mathrm{SE}$, the total zero-delay coincidence is the sum of the three channels already derived [Eqs.~\eqref{eq:chanA}, \eqref{eq:chanB}, \eqref{eq:c0-2SE}], the
emitter--background channel summed over both emitters (total detected mean $\bar\mu$):
\begin{equation}
C_0^{(\mathrm{2SE;1SE+B;2B})}
= C_0^{(\mathrm{2SE})} + C_0^{(\mathrm{1SE+B;2B})}
= N_{\mathrm{cyc}}P_\mathrm{eff}\!
\left[\frac{2r}{(1+r)^2}\bar\mu^{\,2}\mathcal{S}_\mathrm{SE}
+\left(2\bar\mu\bar\beta+\bar\beta^{\,2}\right)\mathcal{S}_\mathrm{B}\right].
\label{eq:c0-full}
\end{equation}
Dividing by $C_\infty=N_{\mathrm{cyc}}P_\mathrm{eff}(\bar\mu+\bar\beta)^2$ and using $\sigma$ [Eq.~\eqref{eq:sigma}] together with the identity
$2\sigma(1-\sigma)+(1-\sigma)^2=1-\sigma^2$,
\begin{equation}
\boxed{\;
g^{(2)}_{\mathrm{meas}}(0)
=\underbrace{\frac{2r}{(1+r)^2}\,\mathcal{S}_\mathrm{SE}}
 _{\displaystyle g^{(2)}_{\mathrm{2SE}}(0)}\;\sigma^2
\;+\;\mathcal{S}_\mathrm{B}\,\left(1-\sigma^2\right).\;}
\label{eq:g2-meas}
\end{equation}
The measured value is a weighted mean of the dead-time-corrected emitter value (weight $\sigma^2$, the squared signal purity) and the background level $\mathcal{S}_\mathrm{B}\approx1$ (weight $1-\sigma^2$); background pulls the measurement toward unity.

\paragraph{Limiting cases.}
Perfect single emitter ($r\to0$):
$g^{(2)}_{\mathrm{meas}}(0)=\mathcal{S}_\mathrm{B}(1-\sigma^2)=(2\rho+1)\mathcal{S}_\mathrm{B}/(\rho+1)^2$; no background ($\sigma\to1$):$g^{(2)}_{\mathrm{meas}}(0)=g^{(2)}_{\mathrm{2SE}}(0)$; pure background ($\sigma\to0$): $g^{(2)}_{\mathrm{meas}}(0)=\mathcal{S}_\mathrm{B}\to1$.

\paragraph{Uncorrected threshold and discrimination margin.}
Without background correction, the operative two-equal-emitter threshold is the $r=1$ value of Eq.~\eqref{eq:g2-meas},
\begin{equation}
g^{(2)}_{\mathrm{thr,meas}}(\sigma)
=\tfrac12\,\mathcal{S}_\mathrm{SE}\,\sigma^2
+\mathcal{S}_\mathrm{B}\,\left(1-\sigma^2\right).
\label{eq:thr-meas}
\end{equation}
The \emph{discrimination margin} is the gap on the $g^{(2)}(0)$ axis between the reading a perfect single emitter produces and that of two equal emitters at the same background level,
\begin{equation}
\Delta g
\equiv g^{(2)}_{\mathrm{thr,meas}}(\sigma)
 - g^{(2)}_{\mathrm{meas}}(0)\big|_{r=0}
= \tfrac12\,\mathcal{S}_\mathrm{SE}\,\sigma^2 .
\label{eq:margin}
\end{equation}
It is suppressed by both dead time ($\mathcal{S}_\mathrm{SE}<1$) and background ($\sigma^2<1$), each bringing the single- and two-emitter readings closer and making contamination harder to distinguish; the common background floor $\mathcal{S}_\mathrm{B}(1-\sigma^2)$ cancels in the
difference.

\paragraph{Background correction.}
Inverting Eq.~\eqref{eq:g2-meas} and renormalizing so that the corrected baseline is unity,
\begin{equation}
g^{(2)}_\mathrm{corr}(0)
=\frac{g^{(2)}_\mathrm{meas}(0)-\mathcal{S}_\mathrm{B}(1-\sigma^2)}{\mathcal{R}_\mathrm{B}},
\qquad
\mathcal{R}_\mathrm{B}\equiv 1-\mathcal{S}_\mathrm{B}(1-\sigma^2),
\label{eq:g2-corr}
\end{equation}
 and equivalently,
\begin{equation}
\boxed{\;
g^{(2)}_\mathrm{corr}(0)
=\frac{2r}{(1+r)^2}\,\mathcal{S}_\mathrm{SE}\,\mathcal{F}_\mathrm{B},
\qquad
\mathcal{F}_\mathrm{B}\equiv\frac{\sigma^2}{\mathcal{R}_\mathrm{B}}.
\;}
\label{eq:g2-corr-2}
\end{equation}
The corrected threshold is therefore
\begin{equation}
\boxed{\;
g^{(2)}_\mathrm{thr,corr}=\tfrac12\,\mathcal{S}_\mathrm{SE}\,\mathcal{F}_\mathrm{B}.\;}
\label{eq:thr-corr}
\end{equation}
With $\rho=\bar\mu/\bar\beta$ and $\mathcal{S}_\mathrm{B}\to1$, Eq.~\eqref{eq:g2-corr-2} reduces to the familiar form
\begin{equation}
g^{(2)}_\mathrm{corr}(0)
=\frac{(1+\rho)^2\,g^{(2)}_\mathrm{meas}(0)-(1+2\rho)}{\rho^2}
=\frac{g^{(2)}_\mathrm{meas}(0)-(1-\sigma^2)}{\sigma^2},
\label{eq:g2-corr-classical}
\end{equation}
with two refinements relative to the standard background-corrected expression~\cite{brouri2000paf}: the subtracted background level is $\mathcal{S}_\mathrm{B}=1-2\tau_\mathrm{d}/\tau_\mathrm{col}$ rather than unity, and the surviving benchmark is $\tfrac12\mathcal{S}_\mathrm{SE}\mathcal{F}_\mathrm{B}$ rather than $\tfrac12$, reflecting dead time in both the emitter--emitter term (via $\mathcal{S}_\mathrm{SE}$) and the incomplete background removal (via $\mathcal{F}_\mathrm{B}$). 

\subsubsection{Statistical uncertainty of \texorpdfstring{$g^{(2)}(m)$}{g2(m)}}
\label{SIsec:errors}

The estimator $\hat g^{(2)}(m)=C_m/C_\infty$ is a ratio whose uncertainty has three parts: the error of the numerator $C_m$, of the denominator $C_\infty$, and their covariance---both are built from the same photon stream and are not independent.
Throughout, we exploit that the $N_\mathrm{cyc}$ repump--probe cycles are statistically independent (the repump resets the charge state), so that both $C_m$ and $C_\infty$ are sums of independent per-cycle contributions and all variances follow model-free from the cycle-to-cycle spread.

$C_m=\sum_k n_k n_{k+m}$ is a sum of \emph{products} of occupation numbers, not of counts.
A single count participates in many pairs (overlapping operands), and spectral diffusion (Sec.~\ref{SIsubsec:add_g2}) imprints super-Poissonian intensity fluctuations; both make $\mathrm{Var}(C_m)>C_m$, so assuming $\mathrm{Var}(C_m)=C_m$ \emph{underestimates} the error.
Writing $C_m=\sum_c c_m^{(c)}$ with $c_m^{(c)}=\sum_{k\in\mathrm{int}}n_{c,k}n_{c,k+m}$ (and $c_0^{(c)}=\sum_{k\in\mathrm{int}}n_{c,k}(n_{c,k}-1)$), independence of cycles gives exactly
\begin{equation}
\mathrm{Var}(C_m)=N_\mathrm{cyc}\,\mathrm{Var}\!\left(c_m^{(c)}\right)
=\frac{N_\mathrm{cyc}}{N_\mathrm{cyc}-1}\sum_c\left(c_m^{(c)}-\bar c_m\right)^2,
\qquad \bar c_m=\frac{C_m}{N_\mathrm{cyc}},
\label{eq:var-cm}
\end{equation}
capturing overlapping-pair correlations, spectral diffusion, and brightness spread with no distributional assumption.
The empirical dispersion $\mathrm{Var}(C_m)/C_m$ measures the departure from Poisson; in our data $\mathrm{Var}(C_0)/C_0\approx2$, i.e.\ genuinely super-Poissonian.

The uncorrelated level $C_\infty=\sum_c t_c$ [Eq.~\eqref{eq:tail}] is likewise a sum of independent per-cycle terms $t_c$.
Its variance and standard error are
\begin{equation}
\mathrm{Var}(C_\infty)=N_\mathrm{cyc}\,\mathrm{Var}(t_c),
\qquad
\mathrm{SEM}(C_\infty)=\sqrt{N_\mathrm{cyc}}\;\mathrm{SD}(t_c),
\quad
\mathrm{SD}(t_c)=\sqrt{\frac{1}{N_\mathrm{cyc}-1}\sum_c\left(t_c-\bar t\,\right)^2},
\label{eq:cinf-sem}
\end{equation}
with $\bar t=C_\infty/N_\mathrm{cyc}$.
The factor $\sqrt{N_\mathrm{cyc}}$ (rather than $1/\sqrt{N_\mathrm{cyc}}$) arises because $C_\infty$ is a \emph{sum}, not a mean, of $N_\mathrm{cyc}$ independent terms, so the variances add: $\mathrm{Var}(C_\infty)=\sum_c\mathrm{Var}(t_c)=N_\mathrm{cyc}\mathrm{Var}(t)$.

Because $C_m$ and $C_\infty$ share the same cycles,
\begin{equation}
\mathrm{Cov}(C_m,C_\infty)=N_\mathrm{cyc}\,\mathrm{Cov}\!\left(c_m^{(c)},t_c\right)
=\frac{N_\mathrm{cyc}}{N_\mathrm{cyc}-1}\sum_c\left(c_m^{(c)}-\bar c_m\right)\left(t_c-\bar t\,\right),
\end{equation}
which is positive (a bright cycle inflates both) and \emph{reduces} the ratio variance.
The delta-method variance is
\begin{equation}
\boxed{\;
\left(\frac{\sigma_{g}}{g}\right)^2
=\frac{\mathrm{Var}(C_m)}{C_m^2}
+\frac{\mathrm{Var}(C_\infty)}{C_\infty^2}
-2\,\frac{\mathrm{Cov}(C_m,C_\infty)}{C_m C_\infty}.\;}
\label{eq:ratio-var}
\end{equation}
All terms follow from the per-cycle series $\{c_m^{(c)}\}$, $\{t_c\}$.
In the high-purity regime the covariance term is negligible.


\paragraph{Uncertainties of the calibration inputs.}
The dead time- and background-corrected estimators depend on four measurable quantities: the effective dead time $\tau_\mathrm{d}$, the emission lifetime $\tau_\mathrm{SE}$, the mean detected signal per window $\bar\nu=\bar\mu+\bar\beta$, and the mean background per window $\bar\beta$.
The dead time $\tau_\mathrm{d}$ is not measured here but taken from the detector datasheet, which specifies it as an upper bound $\tau_\mathrm{d}<\tau_\mathrm{d}^{\max}$.
We use $\tau_\mathrm{d}=\tau_\mathrm{d}^{\max}$ as the most conservative value and assign it no uncertainty. 
The remaining three inputs are extracted from the same photon stream that yields $C_m$: $\bar\nu$ from the interior-pulse count average with cycle-to-cycle standard error $\sigma_{\bar\nu}$, and both $\tau_\mathrm{SE}$ and $\bar\beta$ from a joint fit of the intra-window arrival-time histogram to an exponential decay on a uniform background, with fit errors $\sigma_{\tau_\mathrm{SE}}$ and $\sigma_{\bar\beta}$ (and, in principle, a fit covariance $\mathrm{Cov}(\tau_\mathrm{SE},\bar\beta)$ that we neglect here as it does not enter the derived quantities $\mathcal{S}_\mathrm{SE}$ and $\sigma$ jointly).
We propagate each into the derived quantities $\mathcal{S}_\mathrm{SE}$ and $\sigma$ before combining them into the error on $g^{(2)}_\mathrm{corr}(0)$ and the various reference lines.

\emph{Dead-time survival factor} $\mathcal{S}_\mathrm{SE}=e^{-\tau_\mathrm{d}/\tau_\mathrm{SE}}$ [Eq.~\eqref{eq:RSE-inf}]. With $\tau_\mathrm{d}$ taken as exact, the only remaining input is the lifetime,
\begin{equation}
\frac{\sigma_{\mathcal{S}_\mathrm{SE}}}{\mathcal{S}_\mathrm{SE}}
=\frac{\tau_\mathrm{d}}{\tau_\mathrm{SE}}\,\frac{\sigma_{\tau_\mathrm{SE}}}{\tau_\mathrm{SE}}
=\frac{\tau_\mathrm{d}\,\sigma_{\tau_\mathrm{SE}}}{\tau_\mathrm{SE}^{\,2}}.
\label{eq:sig-SSE}
\end{equation}

\emph{Background survival factor} $\mathcal{S}_\mathrm{B}=1-2\tau_\mathrm{d}/\tau_\mathrm{col}+\mathcal{O}(\tau_\mathrm{d}^2/\tau_\mathrm{col}^2)$ [Eq.~\eqref{eq:Sleading}]. Both $\tau_\mathrm{d}$ and $\tau_\mathrm{col}$ are taken as exact, so
\begin{equation}
\sigma_{\mathcal{S}_\mathrm{B}}=0
\label{eq:sig-SB}
\end{equation}
and $\mathcal{S}_\mathrm{B}$ acts as a fixed numerical constant in every propagation below.

\emph{Signal fraction} $\sigma=1-\bar\beta/\bar\nu$ [Eq.~\eqref{eq:sigma}]. With $\partial\sigma/\partial\bar\nu=\bar\beta/\bar\nu^{\,2}$ and $\partial\sigma/\partial\bar\beta=-1/\bar\nu$,
\begin{equation}
\sigma_\sigma^{\,2}
=\left(\frac{\bar\beta}{\bar\nu^{\,2}}\right)^{\!2}\sigma_{\bar\nu}^{\,2}
+\left(\frac{1}{\bar\nu}\right)^{\!2}\sigma_{\bar\beta}^{\,2}
=\frac{1}{\bar\nu^{\,2}}\left[(1-\sigma)^2\,\sigma_{\bar\nu}^{\,2}+\sigma_{\bar\beta}^{\,2}\right].
\label{eq:sig-sigma}
\end{equation}

\paragraph{Propagation to $g^{(2)}_\mathrm{corr}(0)$.}
From Eq.~\eqref{eq:g2-corr} and using $1-g_\mathrm{meas}=\mathcal{R}_\mathrm{B}(1-g_\mathrm{corr})$, the surviving partial derivatives are
\begin{equation}
\frac{\partial g_\mathrm{corr}}{\partial g_\mathrm{meas}}=\frac{1}{\mathcal{R}_\mathrm{B}},
\qquad
\frac{\partial g_\mathrm{corr}}{\partial\sigma}
=\frac{2\sigma\,\mathcal{S}_\mathrm{B}\,(1-g_\mathrm{corr})}{\mathcal{R}_\mathrm{B}}.
\end{equation}
Treating $g_\mathrm{meas}$ and $\sigma$ as independent inputs (the correlation between them through the shared $\bar\nu$ is subleading in the high-purity regime and is omitted),
\begin{equation}
\boxed{\;
\sigma_{g_\mathrm{corr}}^{\,2}
=\frac{\sigma_{g_\mathrm{meas}}^{\,2}}{\mathcal{R}_\mathrm{B}^{\,2}}
+\frac{4\sigma^{\,2}\mathcal{S}_\mathrm{B}^{\,2}(1-g_\mathrm{corr})^2}{\mathcal{R}_\mathrm{B}^{\,2}}\,\sigma_\sigma^{\,2}.\;}
\label{eq:gcorr-err}
\end{equation}

\paragraph{Propagation to reference lines on the normalized axis.}
The measured correlation and every reference line are drawn on the normalized axis $y=C_m/C_\infty$, so they all share the same denominator $C_\infty$ and inherit its fractional uncertainty
\begin{equation}
\phi\equiv\frac{\mathrm{SEM}(C_\infty)}{C_\infty}
=\frac{\mathrm{SD}(t_c)}{\sqrt{N_\mathrm{cyc}}\;\bar t}
\qquad[\text{Eq.~\eqref{eq:cinf-sem}}].
\label{eq:phi}
\end{equation}
For the data points, $\phi$ is already contained in $\sigma_{g_\mathrm{meas}}$ [Eq.~\eqref{eq:ratio-var}], where it enters together with $\mathrm{Var}(C_m)$ and $\mathrm{Cov}(C_m,C_\infty)$. 
For a reference line at height $y$, which is placed on the axis by dividing a physical coincidence level $y\,C_\infty$ by the measured $C_\infty$, the calibration-input variance must include a $y^{\,2}\phi^{\,2}$ term. 
Below we build up the four horizontal lines encountered in the analysis.

\emph{(i) Uncorrelated level} $y_1=1$. By construction there are no calibration inputs and only the $C_\infty$ error contributes,
\begin{equation}
\sigma_{y_1}^{\,2}=\phi^{\,2}.
\label{eq:err-y1}
\end{equation}

\emph{(ii) Uncorrected two-equal-emitter threshold} $y_\mathrm{thr,meas}=\tfrac12\mathcal{S}_\mathrm{SE}\sigma^{\,2}+\mathcal{S}_\mathrm{B}(1-\sigma^{\,2})$ [Eq.~\eqref{eq:thr-meas}]. 
The calibration-input partials are $\partial/\partial\mathcal{S}_\mathrm{SE}=\tfrac12\sigma^{\,2}$ and $\partial/\partial\sigma=\sigma(\mathcal{S}_\mathrm{SE}-2\mathcal{S}_\mathrm{B})$, giving
\begin{equation}
\sigma_{y_\mathrm{thr,meas}}^{\,2}
=\left(\tfrac12\sigma^{\,2}\right)^{2}\sigma_{\mathcal{S}_\mathrm{SE}}^{\,2}
+\sigma^{\,2}\left(\mathcal{S}_\mathrm{SE}-2\mathcal{S}_\mathrm{B}\right)^{2}\sigma_\sigma^{\,2}
+y_\mathrm{thr,meas}^{\,2}\,\phi^{\,2}.
\label{eq:err-thr-meas}
\end{equation}

\emph{(iii) Background floor} $y_\mathrm{bg}=\mathcal{S}_\mathrm{B}(1-\sigma^{\,2})$. 
This is the level a perfect single emitter reads before subtraction. 
With $\mathcal{S}_\mathrm{B}$ exact, only $\sigma$ and $\phi$ contribute,
\begin{equation}
\sigma_{y_\mathrm{bg}}^{\,2}
=\left(2\mathcal{S}_\mathrm{B}\sigma\right)^{2}\sigma_\sigma^{\,2}
+y_\mathrm{bg}^{\,2}\,\phi^{\,2}.
\label{eq:err-ybg}
\end{equation}

\emph{(iv) Corrected two-equal-emitter threshold} $y_\mathrm{thr,corr}=\tfrac12\mathcal{S}_\mathrm{SE}\mathcal{F}_\mathrm{B}$ with $\mathcal{F}_\mathrm{B}=\sigma^{\,2}/\mathcal{R}_\mathrm{B}$ [Eq.~\eqref{eq:thr-corr}]. 
The only $\mathcal{F}_\mathrm{B}$ partial that contributes is
\begin{equation}
\frac{\partial\mathcal{F}_\mathrm{B}}{\partial\sigma}
=\frac{2\sigma\,(1-\mathcal{S}_\mathrm{B})}{\mathcal{R}_\mathrm{B}^{\,2}},
\end{equation}
and
\begin{equation}
\sigma_{y_\mathrm{thr,corr}}^{\,2}
=\left(\tfrac12\mathcal{F}_\mathrm{B}\right)^{2}\sigma_{\mathcal{S}_\mathrm{SE}}^{\,2}
+\left(\tfrac12\mathcal{S}_\mathrm{SE}\,\frac{\partial\mathcal{F}_\mathrm{B}}{\partial\sigma}\right)^{\!2}\sigma_\sigma^{\,2}
+y_\mathrm{thr,corr}^{\,2}\,\phi^{\,2}.
\label{eq:err-thr-corr}
\end{equation}

\subsection{Additional autocorrelation characterization}
\label{SIsubsec:add_g2}
\begin{figure}[ht]
    \centering
    \includegraphics[width=0.8\linewidth]{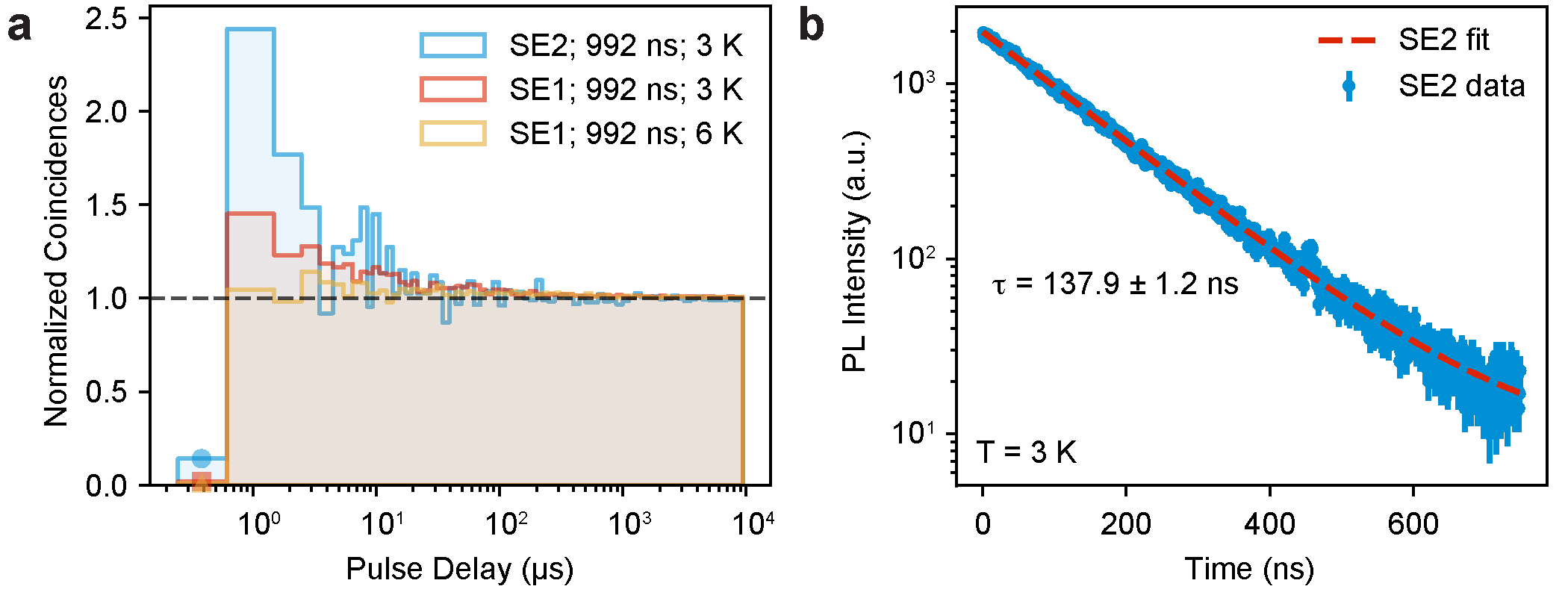}
    \caption{ 
    For each of the following, in each Charge-Repump Sequence we probe 2.5 million times  for 50\,ns, collect for 800\,ns, with a total cycle time of 992\,ns. 
    The repump power was 4.7\,\uW\ and duration was 1\,s.
    \textbf{(a)} Pulsed autocorrelation measurements for isolated peak SE2 and SE1 under at two different temperatures. These plots only show the positive arm in a logarithmic scale on the time axis with logarithmically spaced bins. The 0th bin has been limited to a smaller width but with a marker on its center for visual clarity.
    \textbf{(b)} Optical lifetime measurement of SE2 when on resonance with the cavity showing a reduced lifetime relative to bulk emitters due to Purcell enhancement.
    }
    \label{figS:g2-lifetime-2}
\end{figure}

To investigate the origin of the bunching feature in all $g^{(2)}$ measurements taken at 3\,K, we increased the temperature of the sample to 6\,K and repeated the $g^{(2)}$ characterization for SE1. 
As can be seen from the comparison of the red and green histograms in Figure~\ref{figS:g2-lifetime-2}a, the strong bunching feature at short pulse delays disappears upon the increase in temperature even though the emitter is still bright.
This suggests that the bunching may be a result of time correlations from spectral diffusion. 
Under this model, at longer time scales, spectral diffusion would result in a chance of the defect not being on resonance with the laser during both time bins, reducing the chance of a coincidence for these time delays.
However, at time scales shorter than the spectral diffusion rate, the defect would be more likely to still be on-resonance with the laser following a successful excitation and emission event, thereby creating the bunching.
The disappearance of the bunching at higher temperatures can be explained by this model as a temperature-induced reduction in the characteristic time scale of spectral diffusion to the point where the chance of the defect being on resonance, even in adjacent bins, is no longer correlated and has been reported in other single emitter systems~\cite{dibos2018ass}.

We also performed performed a $g^{(2)}$ measurement and associated lifetime measurement on a fainter isolated peak denoted SE2 (located at $\lambda = 1279.02$\,nm in Figure~\ref{fig3:charge}a of the main text). The $g^{(2)}$ measurement is shown in the Figure~\ref{figS:g2-lifetime-2}a (blue histogram) and confirms that SE2 is also a single emitter with an even more pronounced bunching amplitude at 3 K. However, the associated lifetime measurement (Figure~\ref{figS:g2-lifetime-2}b) shows a slower Purcell-enhanced lifetime likely because the cavity-to-emitter coupling was less optimal than SE1 presented in the main text. 

\printbibliography